\documentclass[12pt,a4paper]{article}
\input epsf
\usepackage{amsmath,amsfonts,epsfig,color,latexsym}
\usepackage{amssymb}
\usepackage[english, USenglish]{babel}
\usepackage{epsfig}
\usepackage{a4wide}
\usepackage{rotating}
\usepackage{float}

\catcode`\@=11
\newcount\hour
\newcount\minute
\newtoks\amorpm
\hour=\time\divide\hour by60
\minute=\time{\multiply\hour by60 \global\advance\minute by-\hour}
\edef\standardtime{{\ifnum\hour<12 \global\amorpm={am}%
        \else\global\amorpm={pm}\advance\hour by-12 \fi
        \ifnum\hour=0 \hour=12 \fi
        \number\hour:\ifnum\minute<10 0\fi\number\minute\the\amorpm}}
\edef\militarytime{\number\hour:\ifnum\minute<10 0\fi\number\minute}
\def\draftlabel#1{{\@bsphack\if@filesw {\let\thepage\relax
   \xdef\@gtempa{\write\@auxout{\string
      \newlabel{#1}{{\@currentlabel}{\thepage}}}}}\@gtempa
   \if@nobreak \ifvmode\nobreak\fi\fi\fi\@esphack}
        \gdef\@eqnlabel{#1}}
\def\@eqnlabel{}
\def\@vacuum{}
\def\draftmarginnote#1{\marginpar{\raggedright\scriptsize\tt#1}}
\def\draft{\oddsidemargin -.2truein
        \def\@oddfoot{\sl Preliminary Notes \hfil
        \rm\thepage\hfil\sl\today\quad\militarytime}
        \let\@evenfoot\@oddfoot \overfullrule 3pt
        \let\label=\draftlabel
        \let\marginnote=\draftmarginnote
   \def\@eqnnum{(\theequation)\rlap{\kern\marginparsep\tt\@eqnlabel}%
\global\let\@eqnlabel\@vacuum}  }
\relax

\newcommand{\be}{\begin{equation}}
\newcommand{\ee}{\end{equation}}
\newcommand{\bea}{\begin{eqnarray}}
\newcommand{\eea}{\end{eqnarray}}
\newcommand{\nn}{\nonumber}

\newcommand{\cN}{{\cal N}}

\newcommand{\bt}[1]{{\bar t}}

\newcommand{\mc}{\mathcal}

\newcommand{\f}{\frac}

\newcommand{\btau}{\bar{\tau}}
\newcommand{\hbe}{\hat{\bar{\mathcal{E}}}_2}
\newcommand{\bor}{\mathcal{G}}
\newcommand{\boru}{\mathcal{G}_{\rm ext}}

\author{\\[-0.3cm]\large Matthias~R.~Gaberdiel\footnote{\tt
gaberdiel@itp.phys.ethz.ch}~, Stefan Hohenegger\footnote{{\tt
shoheneg@mppmu.mpg.de}}~, Daniel~Persson\footnote{{\tt
daniel.persson@itp.phys.ethz.ch}}\,\,\,\footnote{Also affiliated with {\it Fundamental Physics, Chalmers University of Technology, SE-412 96 Gothenburg, Sweden.}}}
\title{\begin{flushright}{\vspace{-0.8cm}\small MPP-2011-7}\end{flushright}
\vspace{0.8cm}Borcherds Algebras and $\cN=4$ Topological Amplitudes}
\dateUSenglish
\date{}
\graphicspath{{figures/}}

\begin{document}

\begin{titlepage}

\maketitle
\begin{center}
\renewcommand{\thefootnote}{\fnsymbol{footnote}}\vspace{-0.5cm}
\footnotemark[1]\footnotemark[2]\footnotemark[3]Institut f\"ur Theoretische
Physik, ETH
  Z\"urich, \\
CH-8093 Z\"urich, Switzerland\\[1cm]
\vspace{-0.3cm}
\footnotemark[2]Max--Planck--Institut f\"ur
Physik\\Werner--Heisenberg--Institut\\
F\"ohringer Ring 6, 80805 M\"unchen, Germany\\[0.5cm]
\end{center}
\begin{abstract}
\noindent 
The perturbative spectrum of BPS-states in the $E_8\times E_8$ heterotic string theory 
compactified on $\mathbb{T}^2$ is analysed. We show that the
space of BPS-states forms a representation of a certain Borcherds algebra
$\bor$ which we construct explicitly using an auxiliary conformal field theory. 
The denominator formula of an extension $\boru \supset \bor$ of this algebra 
is then found to appear in a certain heterotic one-loop $\mathcal{N}=4$ topological
string amplitude. Our construction thus gives an $\mathcal{N}=4$
realisation of the idea envisioned by Harvey and Moore, namely that the `algebra
of BPS-states' controls the threshold corrections in the heterotic string.
\end{abstract}

\end{titlepage}
\noindent

\newpage 
\tableofcontents

\renewcommand{\theequation}{\arabic{section}.\arabic{equation}}
\section{Introduction}

The study of BPS-states has played a prominent role in developing our current
understanding 
of string theory. Quantities which only receive contributions from BPS-states
(`threshold corrections') 
are protected under variations of the string coupling and are therefore ideal
for probing strong-weak 
dualities and non-perturbative effects in string theory. Degeneracies of
BPS-states (or rather BPS-indices) 
are moreover closely related to interesting mathematical structures, such as the
counting of 
rational curves and special Lagrangian submanifolds in Calabi-Yau manifolds, or
(generalised) Donaldson-Thomas invariants. The BPS-index is locally constant as
a function 
of the moduli, but may jump at codimension one submanifolds, known as walls of
marginal stability, 
on which bound states of BPS-states may decay or recombine
\cite{Cecotti:1992rm,Seiberg:1994rs}. 
The behaviour of the BPS-index when crossing such walls has been the subject of
intense research (see, 
{\it e.g.}\
\cite{Sen:2007vb,Denef:2007vg,Cheng:2008fc,Gaiotto:2008cd,js,ks,
Alexandrov:2008gh,Manschot:2010qz}),
leading to interesting wall-crossing formulae with a broad range of applications
in mathematics and physics. 

In two insightful papers \cite{Harvey:1995fq,Harvey:1996gc} (see also
\cite{Moore:1997ar} for a nice overview), Harvey and Moore argued
that the space of BPS-states in string theory forms an algebra, and they
provided evidence that this 
`algebra of BPS-states' should be related to a (generalised) Borcherds-Kac-Moody
(BKM) algebra 
\cite{Borcherds0}.\footnote{The theory of (super-) BKM-algebras has made its
appearance 
in many different corners of string theory in the past (see, {\it e.g.}\ 
\cite{Gebert:1994mv,Barwald:1997gm,Gebert:1997hx,HenryLabordere:2002dk,%
HenryLabordere:2002xh,Henneaux:2010ys}).} In particular, they found that
certain 
threshold corrections in heterotic $\mathcal{N}=2$ compactifications can be
written as 
infinite product representations of automorphic forms on the Grassmannian 
$SO(2,2+n)/(SO(2)\times SO(2+n))$, which, through the work of Borcherds 
\cite{Borcherds1}, are in turn related to denominator formulae for
BKM-algebras. 
Although these results are intriguing and suggestive, a direct connection
between these 
infinite product formulae and the algebra of BPS-states has not yet been
established.

In the context of type II Calabi-Yau compactifications, the algebra of
BPS-states was further 
analysed in \cite{Neumann:1997pr,Fiol:2000wx}. In \cite{Neumann:1997pr} the
vertex 
algebra realisation of the BPS-algebra was developed more explicitly, and it
was, in 
particular, shown how a BKM-algebra appears as a certain subalgebra of the full
algebra 
of BPS-states. A different point of view was taken in \cite{Fiol:2000wx}, where
the description of 
D-brane states in terms of quivers was exploited. In this context, the algebra
of BPS-states was 
related to quiver representations, and in certain specific examples this
analysis revealed the 
BPS-algebra appearing as an affine Kac-Moody algebra attached to the D-brane
quiver. 
However, the relation between the algebra of BPS-states and threshold
corrections 
remained obscure. Recently it has also been suggested that the correct
mathematical 
framework for analysing the algebra of BPS-states in Calabi-Yau
compactifications is 
through the so called `cohomological Hall algebra' \cite{ks2}. 

A seemingly different development started with the work of 
Dijkgraaf, Verlinde and Verlinde \cite{Dijkgraaf:1996it}, who conjectured that
the degeneracies 
of (non-perturbative) dyonic $1/4$ BPS-states in $\mathcal{N}=4$ heterotic
compactifications 
are encoded in the Fourier coefficients of a certain Siegel modular form, known
as the 
`Igusa cusp form'. As had earlier been shown by Gritsenko and Nikulin
\cite{GritsenkoNikulin,Gritsenko:1996ax}, 
this Igusa cusp form has an infinite product representation which relates it to
the denominator 
formula of a certain rank 3 (super) BKM-algebra, denoted $\mathfrak{g}_{{\rm
1,II}}$ in \cite{GritsenkoNikulin}. 
In this way, the degeneracies of dyons also become related to the root
multiplicities of the associated 
BKM-algebra. The physical role of the algebra $\mathfrak{g}_{{\rm 1,II}}$ was
further clarified in 
\cite{Cheng:2008fc} (see also
\cite{Cheng:2008gx,Cheng:2008kt,Govindarajan:2008vi,Govindarajan:2009qt,Govindarajan:2010fu}), 
where it was shown that the wall-crossing behaviour of the dyon spectrum is controlled
by the hyperbolic Weyl group $\mathcal{W}(\mathfrak{g}_{{\rm 1,II}})$ of this BKM-algebra.
\smallskip

The purpose of the present work is to elucidate the relation between the algebra
of  BPS-states and the infinite product formulas occurring in certain BPS couplings
in heterotic string theory. 
Our analysis is inspired by the results of \cite{Harvey:1995fq}, but, as will
become clear below, the approach 
we take is slightly different. We focus our attention on the spectrum of the
$E_8\times E_8$ heterotic 
string compactified on a six-torus ${\mathbb T}^{6}$. For simplicity, we further
assume that the torus 
splits according to ${\mathbb T}^{6}={\mathbb T}^{4}\times {\mathbb T}^2$, and take the large volume 
limit of the ${\mathbb T}^4$. The Narain moduli space of the theory is therefore
given by the arithmetic 
coset $SO(2,18;\mathbb{Z})\backslash SO(2,18)/(SO(2) \times SO(18))$, where
$SO(2,18;\mathbb{Z})$ is
the U-duality group which leaves the lattice of BPS-charges invariant. The
spectrum of perturbative 
BPS-states corresponds to taking the right-moving sector of the heterotic
string to be in its ground state, 
while allowing for arbitrary excitations in the left-moving sector. We show
that the space of BPS-states
forms a representation of a certain BKM-algebra $\bor$, which we explicitly
construct using
an auxiliary bosonic conformal field theory. The BKM-algebra has root lattice 
$\Pi^{1,1}\oplus \Lambda_{\mathfrak{e}_8}\oplus \Lambda_{\mathfrak{e}_8}$ 
and can therefore be understood as a Borcherds extension of the Lorentzian 
Kac-Moody algebra $(\mathfrak{e}_8\oplus \mathfrak{e}_8)^{++}$. 
 
The auxiliary CFT construction actually leads to a slightly bigger Borcherds 
algebra $\boru$, that contains $\bor$ as a natural subalgebra, 
$\bor\subset\boru$. The extended algebra $\boru$ is 
based on the root lattice $\Pi^{1,1}\oplus \Lambda_{\mathfrak{e}_8}\oplus 
\Lambda_{\mathfrak{e}_8}\oplus \Lambda_{\mathfrak{e}_8}$. While it does not 
directly  act on the space of BPS-states, it turns out to be relevant for understanding 
the algebraic structure of threshold corrections. 
To make this precise, we consider 
a particular class of BPS-saturated $g$-loop amplitudes $\mathcal{F}_g$ in  type II 
string theory compactified  on K3$\times {\mathbb T}^2$ which are captured by correlation 
 functions of the $\mathcal{N}=4$ topological string \cite{Antoniadis:2006mr}.
For any $g$, the dual 
 amplitudes in heterotic string theory compactified on ${\mathbb T}^6$ receive
contributions at all loop 
 orders in (heterotic) perturbation theory. However, the leading contribution in
the heterotic weak coupling 
 limit is a one-loop expression  which is therefore amenable to an analysis
similar to that of 
 Harvey and Moore in the $\mathcal{N}=2$ setting
\cite{Harvey:1995fq,Harvey:1996gc}. Mathematically, 
 heterotic one-loop amplitudes fall into the category of so called `singular
theta correspondences', 
 as analysed in detail by Borcherds \cite{Borcherds1,Borcherds2}. 

As a consequence of the 1/2 BPS nature of the corresponding effective couplings
in supergravity, 
supersymmetric Ward identities predict that $\mathcal{F}_g$ satisfies particular
second order differential 
equations, refered to as \emph{harmonicity equations}
\cite{Antoniadis:2006mr,Antoniadis:2007cw} 
(see also \cite{Berkovits:1994vy,Ooguri:1991fp,Antoniadis:2007ta}). In string theory these
equations get modified by 
anomalous world-sheet boundary terms which signal non-analytic contributions at
the 
quantum level. This modification can thus be understood as an $\mathcal{N}=4$
analogue of the 
holomorphic anomaly of the $\mathcal{N}=2$ B-model topological string 
\cite{Bershadsky:1993ta,Bershadsky:1993cx}. A peculiar feature of the amplitudes
$\mathcal{F}_g$ 
that appear in our context is that one can isolate an anomaly free part
$\mathcal{F}_g^{\text{analy}}$ 
which is analytic and plays an analogous role as the 
threshold corrections of \cite{Harvey:1995fq,Harvey:1996gc}. We evaluate the
integral 
$\mathcal{F}_g^{\text{analy}}$ explicitly using the method of orbits
\cite{Dixon:1990pc,Harvey:1995fq} 
(or in mathematical parlance, the Rankin-Selberg method \cite{Bump}). 

In order to make contact with the BKM-algebras $\bor$ and  $\boru$ discussed above, 
we first analyse the complex codimension one submanifolds 
where the integral $\mathcal{F}_g^{\text{analy}}$ develops singularities. They include 
the walls of the fundamental Weyl chamber of the double extension 
$(\mathfrak{e}_8\oplus \mathfrak{e}_8)^{++}$.\footnote{We note that similar structures 
related to hyperbolic Kac-Moody algebras also play a crucial role in a very different 
physical situation, namely the study of gravity close to a cosmological singularity 
\cite{Damour:2000hv,Damour:2001sa,Damour:2002et,Henneaux:2007ej}. See 
appendix \ref{App:ExtendSemiSimple} for a review of extensions of semisimple Lie
algebras.} This
is similar to what was found for the non-perturbative $1/4$ BPS dyon spectrum 
in \cite{Cheng:2008fc}. Hence, the Weyl group of the Borcherds algebra $\bor$ controls the singularity 
structure of $\mathcal{F}_1^{\text{analy}}$.
We further show that the analytic integral $\mathcal{F}_1^{\text{analy}}$ can
be written as a specialised denominator identity based on the algebra $\boru$, 
where the Cartan angles corresponding to the extra $\mathfrak{e}_8$ are set to zero. 
Our construction thus
gives an explicit $\mathcal{N}=4$ realisation of the idea envisioned by Harvey
and Moore, 
namely that the algebra of BPS-states controls the threshold corrections in the
heterotic string. 
\medskip

The paper is organised as follows. In section \ref{AlgebraBPSstates} we 
discuss aspects of the perturbative BPS-spectrum of the heterotic string on 
${\mathbb T}^6$, and introduce the BKM-algebra $\bor$ and its extension $\boru$,
that will play a key role in the remainder of the paper. In 
section \ref{GeneralAspects} we recall the structure of the $\mathcal{N}=4$
topological amplitudes 
$\mathcal{F}_g$, with emphasis on the role of harmonicity. We show, in
particular, how to single out 
the anomaly free part $\mathcal{F}_g^{\text{analy}}$ of the full amplitude
$\mathcal{F}_g$. In 
section~\ref{SingularityStructure} we restrict the analysis to the large volume
limit of 
${\mathbb T}^4\subset {\mathbb T}^{6}$, and analyse in detail the singularity
structure of the 
resulting integral as a function on the Narain moduli space
$SO(2,18)/(SO(2)\times SO(18))$. In particular, we show that the 
rational quadratic divisors (introduced by Borcherds in \cite{Borcherds1}) coincide 
with the walls of the 
fundamental Weyl chamber of the Lorentzian Kac-Moody algebra
$(\mathfrak{e}_8\oplus \mathfrak{e}_8)^{++}$. 
Section~\ref{Section:Akintegral} is then 
devoted to evaluating the integral $\mathcal{F}_{g=1}^{\text{analy}}$
explicitly. We show 
how to write part of the result in terms of an infinite product, which we
identify with (a restriction of) the denominator 
formula of the BKM-algebra $\boru$. The paper
includes two appendices. 
In appendix~\ref{AppA} we introduce some relevant background on
(in)finite-dimensional 
Lie algebras. In particular, we describe double extensions of both simple and
semisimple finite 
Lie algebras. We also discuss general aspects of Borcherds-Kac-Moody algebras. In 
appendix~\ref{App:Orbits} we collect some details on the calculation of the
integral presented in section~\ref{Section:Akintegral}.

\section{BPS-States and Borcherds-Kac-Moody Algebras}\label{AlgebraBPSstates}

Let us begin by studying the $1/2$ BPS-states in the heterotic theory. Inspired
by \cite{Harvey:1995fq,Harvey:1996gc} (see also \cite{Neumann:1997pr}) we want
to show that they form a 
representation of a certain Borcherds-Kac-Moody (BKM) algebra, which we shall
construct using 
conformal field theory methods. The denominator formula of a closely related BKM algebra will
later play a role in the topological amplitudes that will be calculated in
sections~\ref{Section:HarmonicityGenus1} 
and \ref{Section:Akintegral}. 
In the following we shall consider the $E_8\times E_8$ theory compactified on
${\mathbb T}^6$. 

\subsection{BPS States in Narain Compactifications}
\label{NarainComp}

The classical moduli space for heterotic string theory on ${\mathbb T}^6$ is
described by the coset space
\be\label{classmod}
\mathcal{M}= \Bigl( SL(2,\mathbb{R})/U(1)  \Bigr) \, \times \, \Bigl(
SO(6,22)/(SO(6)\times SO(22))\Bigr)\ ,
\ee
where the first factor encodes the heterotic `axio-dilaton', while the second
factor 
accounts for the Narain moduli of the torus. The perturbative spectrum of the
heterotic string
consists of the states that are created from a momentum ground state labelled
by $(p^L,\vec{p};p^R,\vec{p}\;\! )$ by the action of the oscillators. Here the
compactified (and internal) 
momenta  take values in the Narain lattice
\be
(p^{L}, p^{R}) \in \Gamma^{6,22}\ ,\label{NarainMomenta}
\ee
while $\vec{p}$ describes the momentum in the uncompactified 4-dimensional
theory.
The Narain lattice is invariant under the  T-duality group
$SO(6,22;\mathbb{Z})$, and thus the
quantum moduli space is the quotient of (\ref{classmod}) by this arithmetic
group. 

We shall mainly work in the covariant formulation, where $\vec{p}\in {\mathbb
R}^{3,1}$. The 
physical states of the heterotic string (in the NS sector, say) have to be
annihilated by $L_n$, $n>0$,
and $G_r$, $r>0$, as well as $\bar{L}_n$, $n>0$.\footnote{In our conventions,
the left-movers 
are `supersymmetric', while the right-movers are `bosonic'. 
The right-movers are denoted by a bar.} In addition, they have to satisfy the
level 
matching and mass shell conditions
\begin{equation}\label{levma}
\begin{array}{rcl}
\tfrac{1}{2} & = & \tfrac{1}{2}  (p^L)^2 + \tfrac{1}{2} {\vec{p}}^{\ 2} + N_L\\
1 & = &  \tfrac{1}{2}  (p^R)^2 + \tfrac{1}{2} {\vec{p}}^{\ 2} + N_R \ ,
\end{array}
\end{equation}
where $N_L$ and $N_R$ are the left- and right-moving excitation numbers. The
mass of
a physical state is determined via
\begin{equation}
M^2=(N_L-\tfrac{1}{2}) + \tfrac{1}{2}(p^{L})^2 \ .
\end{equation} 

Since the compactification of 
heterotic string theory on ${\mathbb T}^6$ preserves 
$\mathcal{N}=4$ supersymmetry, the massive \emph{BPS-states} come in two classes: 
$1/2$ BPS-states associated with short multiplets, and
$1/4$ BPS-states associated with intermediate multiplets. As has been 
discussed in \cite{Lerche:1999ju} all $1/4$ BPS-states are non-perturbative, 
and only the $1/2$ BPS-states are perturbative.
For the latter we have in addition to (\ref{levma}) that $N_L=\tfrac{1}{2}$, and
thus the mass is simply 
\begin{equation}
M_{\text{BPS}}^2 = \tfrac{1}{2}(p^{L})^2 \ .
\end{equation} 
For these states we can subtract the two equations in (\ref{levma}) from one
another and obtain
\begin{equation}\label{2.6}
1 = -\tfrac{1}{2} (p^L,p^R)^2 + N_R \ , 
\end{equation}
where $(p^L,p^R)^2 = (p^L)^2 - (p^R)^2$ is the inner product in $\Gamma^{6,22}$.

\subsection{Eight Dimensions}\label{sec:sublattice}

In order to simplify the analysis we shall now consider the decompactification
limit to eight dimensions. This is to say, we split ${\mathbb T}^6={\mathbb T}^2\times {\mathbb T}^4$, 
and take the large-volume limit of the ${\mathbb T}^4$,  effectively setting
the 
${\mathbb T}^4$ momenta to zero. This corresponds to restricting
ourselves to momentum ground states in the even self-dual lattice
$\Gamma^{2,18}$ of signature 
$(2,18)$, where
\begin{equation}
\Gamma^{6,22} = \Gamma^{2,18} \oplus \Gamma^{4,4} \ ,
\end{equation}
and $\Gamma^{4,4}$ describes the momenta of the ${\mathbb T}^4$. 

The elements in $\Gamma^{2,18}$ characterise the momentum ground states of the
heterotic $E_8\times E_8$ string, compactified on ${\mathbb T}^2$. The moduli space of
such compactifications is 
described by the K\"ahler ($T$) and complex ($U$) structure moduli of ${\mathbb
T}^2$,  as well as
by two real Wilson lines $\vec{v}_\alpha\in{\mathbb R}^{16}$, $\alpha=1,2$. At
any point in this 
moduli space, a general element of the momentum lattice
$\Gamma^{2,18}$ can be parametrised as $x=(m_1,n_1;m_2,n_2;\vec{\ell}\,\,)$, where 
$(m_1,m_2)$ and $(n_1,n_2)$ are the momentum and winding numbers along 
${\mathbb T}^2$, while 
$\vec{\ell}\in\Lambda_{\mathfrak{e}_8} \oplus \Lambda_{\mathfrak{e}_8}$. We will also use the notation $\vec{\ell}=(\vec{\ell}_1,\vec{\ell}_2)$ with  $\vec{\ell}_{1,2}\in\Lambda_{\mathfrak{e}_8}$ respectively. The
inner product on $\Gamma^{2,18}$ is defined by 
\begin{align}
\left< x|
x'\right>&=-m_1n_1^{\prime}-n_1m_1^{\prime}-m_2n_2^{\prime}-n_2m_2^{\prime} 
+ \vec{\ell}\cdot \vec{\ell}^{\; \prime}\nonumber\\
&=-m_1n_1^{\prime}-n_1m_1^{\prime}-m_2n_2^{\prime}-n_2m_2^{\prime} 
+ \vec{\ell}_1\cdot \vec{\ell}_1^{\; \prime}+ \vec{\ell}_2\cdot \vec{\ell}_2^{\; \prime}\ ,\label{inpro}
\end{align}
where the first four terms represent the Lorentzian inner product on 
$\Gamma^{2,2}\simeq \Pi^{1,1}\oplus \Pi^{1,1}$, and the last term is the
standard Euclidean inner product inherited from 
${\mathbb R}^{16}\supset \Lambda_{\mathfrak{e}_8} \oplus \Lambda_{\mathfrak{e}_8}$.
For given $x=(m_1,n_1;m_2,n_2;\vec{\ell}\,\, )\in \Gamma^{2,18}$, the actual internal
momentum is then 
a vector in ${\mathbb R}^{16}$
\begin{equation}\label{bdef}
\vec{P}(x) = n_1\, \vec{v}_1 + n_2\, \vec{v}_2 + \vec{\ell} \ .
\end{equation}

For the following it is useful to combine $\vec{v}_1$ and $\vec{v}_2$ into a complex
Wilson line, $\vec{V}= \vec{v}_1 + i \vec{v}_2$. Sometimes we will also use the notation 
$\vec{V}=(\vec{V}_1,\vec{V}_2)$ with $\vec{V}_{1,2}\in \mathbb{C}^8$, respectively.
We parametrise an arbitrary point in the  moduli space by $y=(U,T; \vec{V})\in {\mathbb C}^{1,17}$, 
with inner product 
\begin{align}
(y|y')=-TU'-T'U+\vec{V}\cdot \vec{V}'=-TU'-T'U+
\vec{V}_1\cdot \vec{V}_1'+\vec{V}_2\cdot \vec{V}_2'\,,\label{innerProdRed}
\end{align} 
which particularly implies $(y|y) = - 2\, T\, U + \vec{V}^2 $ for the norm of $y$. 
For the following it is also useful to define the map (see \cite{Harvey:1995fq})
\begin{equation}
u : {\mathbb C}^{1,17} \rightarrow {\mathbb C}^{2,18} \ , \quad 
y=(U,T; \vec{V}) \ \mapsto \ u(y) = \left(U,T;   \frac{(y|y)}{2},1 ;
\vec{V}\right) \ ,
\end{equation}
which associates to every element $y\in {\mathbb C}^{1,17}$ a light-like
vector 
$u(y)\in {\mathbb C}^{2,18}$. Here the inner product on ${\mathbb C}^{2,18}$
is defined
by $\langle \cdot | \cdot \rangle$ as in (\ref{inpro}). With this notation an
arbitrary momentum state $x\in\Gamma^{2,18}$ parametrised by 
$x=(m_1,n_1;m_2,n_2;\vec{\ell}\,\, )$ has 
\begin{align}
|p^L|^2 &= -2 \frac{|\left< x | u(y) \right>|^2}{(\Im y|\Im y)} = 
\frac{1}{\bigl(T_2 U_2-\frac{1}{2}\Im \vec{V}^2\bigr)}\, \Big|m_2+m_1T+n_1U
+\frac{n_2}{2}(y|y) -\vec{\ell}\cdot \vec{V} \Big|^2 
 \nn \\
\Bigl(|p^{R}|^2-|p^{L}|^2\Bigr)&= \left< x | x\right> =\vec{\ell}^{\;
2}-2m_1n_1-2m_2n_2\equiv 2D\ ,\label{discrim}
 \end{align}
where $\Im y=(U_2, T_2; \Im \vec{V})$ is the imaginary part of $y$. 
\subsection{The BKM Algebra}\label{Sect:BKMalgebra}
\label{sec_BKMconstruction}

Next we want to construct a Borcherds-Kac-Moody (BKM) algebra $\bor$
that has a natural action on the space of BPS-states and that will be 
relevant for the topological amplitudes we shall calculate below. This algebra
will arise as a subalgebra of a closely related algebra $\boru$ that we shall construct first.

Let us denote by $\Gamma^{1,17}\subset \Gamma^{2,18}$ the sublattice of the even
self-dual lattice $\Gamma^{2,18}$ which is obtained by setting the 
momenta and windings along the second circle to zero, $m_2=n_2=0$. The lattice
$\Gamma^{1,17}$ is an even self-dual lattice of signature $(1,17)$. We can adjoin 
the even self-dual root lattice $\Lambda_{\mathfrak{e}_8}$ of dimension $8$ to
it, and define the even self-dual lattice $\Gamma^{1,25}$ via
\begin{equation}\label{adde}
\Gamma^{1,25} = \Gamma^{1,17} \oplus \Lambda_{{\mathfrak e}_8}  \ .
\end{equation}
On $\Gamma^{1,25}$ we then consider an (auxiliary) chiral conformal field theory, whose
`physical' states $\psi$ are characterised by the property  that  they are annihilated by 
the Virasoro generators $L_n$ with $n>0$, together with the mass-shell condition that  
the $L_0$ eigenvalue is one, 
\be
L_n\, \psi=0 \quad n>0\ , \qquad \qquad\qquad  L_0\, \psi=\psi\ .
\ee
As is familiar from string theory, these physical states form a Lie algebra, where 
one defines the bracket as \cite{Bor0}
\begin{equation}\label{Lieb}
[\psi , \phi ] = V_0(\psi) \phi \ .
\end{equation}
Since $\psi$ is Virasoro primary and of conformal dimension $h=1$, $[L_n,V_0(\psi)]=0$, 
and thus the right hand side is again a physical state. Since the underlying
momentum space has only one time-like direction, the resulting  Lie algebra is in fact 
a Borcherds-Kac-Moody (BKM) algebra \cite{Bor0}, and we denote it by 
$\boru$.

We can label the elements of $\Gamma^{1,25}$ as 
$(m,n;\vec{\ell}\,\,)$, where $(m,n)$ describes the component in $\Pi^{1,1}$ and 
$\vec{\ell}\in\Lambda_{{\mathfrak e}_8}\oplus \Lambda_{{\mathfrak e}_8}\oplus \Lambda_{{\mathfrak e}_8}$. 
The mass-shell condition in the auxiliary conformal field theory is then 
\begin{equation}\label{auxmass}
1 = N_{\rm exc} + \frac{1}{2} \vec{\ell}^{\ 2} -  m n  \ ,
\end{equation}
where $N_{\rm exc}$ denotes the `excitation' number in the $26$ directions of
the auxiliary chiral conformal field theory. Given $(m,n,\vec{\ell}\,\,)$, this equation can be 
solved for $N_{\rm exc}$, and thus we can directly determine the multiplicity 
$c_{\rm ext}(m,n;\vec{\ell}\,\,)$ of 
the roots corresponding to $(m,n;\vec{\ell}\,\,)$. Their generating function is of the form
\be
\sum_{N\geq -1}\,\,\sum_{\vec{\ell}\in \Lambda_{\mathfrak{e}_8}\oplus
\Lambda_{\mathfrak{e}_8} \oplus\Lambda_{\mathfrak{e}_8}}
c_{\rm ext}(N,\vec{\ell}\,\,) \, \bar{q}^{N}\, 
e^{2\pi i \vec{\ell}\cdot \vec{z}}=
\frac{\Theta_{\mathfrak{e}_8\oplus \mathfrak{e}_8 \oplus \mathfrak{e}_8}(\bar\tau,\vec{z})}{\eta(\bar{q})^{24}}\ ,
\label{GeneralCountingFormula}
\ee
where  $N\equiv mn$ and $\Theta_{\mathfrak{g}}(\tau,\vec{z})$ is the usual theta series of the
root lattice $\Lambda_{\mathfrak g}$. Here $q=e^{2\pi i \tau}$, and 
$\vec{z}=(\vec{z}_1, \vec{z}_2, \vec{z}_3)$ is a $24$-dimensional vector
in the weight space of $\mathfrak{e}_8\oplus \mathfrak{e}_8 \oplus \mathfrak{e}_8$. 
Note that (\ref{CountingFormula}) is a weak Jacobi form, 
and consequently the Fourier coefficients $c_{\rm ext}(N, \vec{\ell}\;)$ only depend on 
$(N, \vec{\ell}\;)$ through the combination 
$N-\vec{\ell}\cdot \vec{\ell} /2$ \cite{EZ}.

In the above construction the additional ${\mathfrak e}_8$ lattice in (\ref{adde})
was introduced by hand, and was not really crucial for the definition; in fact, the construction
would have worked equally for any even lattice of dimension eight.
It is therefore natural to restrict $\boru$ to the subalgebra that
is generated by those physical states for which the momenta actually lie in the sublattice
$\Gamma^{1,17}$; the resulting subalgebra will be denoted by $\bor\subset \boru$. Note that
$\bor$ is indeed a consistent subalgebra since momentum is additive under the product in 
(\ref{Lieb}), and thus the bracket (\ref{Lieb}) closes on $\bor$. The root multiplicities
of $\bor$ are described by 
\be
\sum_{N\geq -1}\,\,\sum_{\vec{\ell}\in \Lambda_{\mathfrak{e}_8}\oplus
\Lambda_{\mathfrak{e}_8}}
c(N,\vec{\ell}\,\,) \, \bar{q}^{N}\, 
e^{2\pi i \vec{\ell}\cdot \vec{z}}=
\frac{\Theta_{\mathfrak{e}_8 \oplus \mathfrak{e}_8}(\bar\tau,\vec{z})}{\eta(\bar{q})^{24}}\ ,
\label{CountingFormula}
\ee
where $\vec{z}=(\vec{z}_1, \vec{z}_2)$ is a $16$-dimensional vector in the weight space of 
$\mathfrak{e}_8\oplus \mathfrak{e}_8$. 

By construction the Lie algebra $\bor$ is a Borcherds
extension of ${\mathfrak g}^{++}$, where ${\mathfrak g}= \mathfrak{e}_8\oplus \mathfrak{e}_8$. Here 
$\mathfrak{g}^{++}$ is the so-called double extension of the finite
dimensional Lie algebra $\mathfrak{g}$, which is a Lorentzian
Kac-Moody algebra with root lattice 
\begin{equation}
\Gamma^{1,17}=\Pi^{1,1}\oplus \Lambda_{\mathfrak{e}_8} \oplus  \Lambda_{\mathfrak{e}_8} \ ,
\label{Lorentziane8e8lattice}
\end{equation}
for more details about double extensions see appendix~\ref{App:DoubleExtension}.
The definition of $\bor$ is similar to the  construction of ${\cal H}_0^{\text{mult}}$ in
\cite{Harvey:1996gc}, see also \cite{Neumann:1997pr}.

To each BKM-algebra one can associate its \emph{denominator formula}. This is obtained by 
restricting the standard Weyl-Kac-Borcherds character formula to the trivial representation, yielding an 
equivalence between an infinite sum over the Weyl group and an infinite product over the positive 
roots (see Appendix \ref{BKMalgebras} for some details). For our purposes it is the infinite product side 
of the denominator formula that will play a crucial role. Let $\Delta_{\boru}$ be the root lattice of $\boru$ and 
denote an arbitrary root by $\alpha$. In this case the infinite product 
part of the general denominator formula (\ref{denominatorformula}) may be written as
\be
\Phi_{\boru}(\hat{y}) = \prod_{\alpha\in \Delta_{\boru}^{+}} 
\left(1-e^{2\pi i (\alpha|\hat{y})}\right)^{c_{\text{ext}}\left(-\alpha^2/2\right)}\ ,
\label{borudenominator}
\ee
where the multiplicity of the root $\alpha$ is given by the Fourier coefficients in 
(\ref{GeneralCountingFormula}) via 
\be
\text{mult}\, \alpha=c_{\text{ext}}\left(-\tfrac{1}{2}\alpha^2\right)
=c_{\text{ext}}\left(mn-\tfrac{1}{2}\vec{\ell}\cdot \vec{\ell}\;\right)\ . 
\ee
The moduli vector $\hat{y}$ is valued in $\mathbb{C}^{1,25}$, and can be chosen as 
$\hat{y}=(y, \vec{z}_3)$, where $y=(U,T; \vec{V})$ represents
the standard Narain moduli and $\vec{z}_3\in \mathbb{C}^8$ is a vector in the 
weight space of the auxiliary $\mathfrak{e}_8$ used in the construction of $\boru$.

\subsection{Action on the BPS-States}

In the following we want to show that the BKM algebra $\bor$ plays a natural
role in the
description of the theory. While it is not in one-to-one correspondence with the
space of BPS-states --- and hence does not deserve the name `BPS algebra' --- it has a 
{\em natural action} on the space of BPS-states. To see this, we simply observe
that 
to each BPS-state with $(m_2,n_2)=(0,0)$ and with no oscillator excitation in
the corresponding
circle direction, we can associate an element $\phi$ of the
auxiliary conformal field theory associated to $\Gamma^{1,25}$. Indeed, we
ignore
the left-moving oscillators (with $N_L=\frac{1}{2}$), and identify $(p^L,p^R)$
with an element $(m_1,n_1,\vec{l}\;)\in\Gamma^{1,17} \subset \Gamma^{1,25}$. 
Furthermore, we identify the $16$ internal right-moving
oscillators, as well as the right-moving oscillator associated to the
$(m_1,n_1)$ direction to 
the oscillators of the auxiliary conformal field theory corresponding to the
$17$ right-moving 
momenta in $\Gamma^{1,17}$. The remaining eight right-moving oscillators of the
BPS-state are 
finally identified with suitable oscillators corresponding to the
$\Lambda_{{\mathfrak e}_8}$ 
lattice of the auxiliary conformal field theory. Because of (\ref{discrim}), 
eq.\ (\ref{2.6}) can then be interpreted 
as the `physical mass-shell condition' (\ref{auxmass}) in the auxiliary conformal field theory.

The Lie bracket (\ref{Lieb}) now defines an action of $\psi\in\bor$ on $\phi$,
and the image under this action is again associated with a BPS-state with the above
properties. Thus we can {\em define} an action of $\bor$ on the space of such BPS-states by letting
$\psi$ act trivially on the left-moving oscillators (which we ignored in mapping the BPS
state to an element of the auxiliary conformal field theory). 

It remains to show that this action can also be extended to the BPS-states for which
$(m_2,n_2)\neq (0,0)$, and that also have oscillators in the corresponding
circle direction. As regards the oscillators, we define the action of $\psi$ to be
trivial on them. (This is consistent since $\psi$ does not carry any momentum along this
direction.) If  $(m_2,n_2)\neq (0,0)$, $(p^L,p^R)$ is no longer an element of $\Gamma^{1,25}$, 
but the action of $\psi$ can still be defined on it since both the
momenta $(p^L,p^R)$ and the momentum associated to $\psi$ are elements of the even
self-dual lattice $\Gamma^{2,18}$. Thus the corresponding fields are local relative to one
another, and the contour integral that is implicit in (\ref{Lieb}) is well-defined. Thus
we can extend the action of $\psi\in\bor$ to all BPS-states, thereby proving our claim. 
\smallskip

We have therefore shown that the BKM algebra $\bor$ acts naturally on the 
{\em full space of perturbative BPS-states}. The algebra itself, however, is only associated to a 
{\em subspace} of BPS-states, namely to those with momentum in $\Gamma^{1,17}$
--- in particular, we required that $(m_2,n_2)=(0,0)$ in order to have only one time-like
direction, leading to a 
BKM algebra. Furthermore, in defining the Lie algebra, we 
ignored the choice of polarisation for the left-movers. (This is similar to
the construction in \cite{Neumann:1997pr}.) As a consequence
our construction does not define a `BPS-algebra', {\it i.e.}\ it does not make
the full space
of BPS-states into a Lie algebra, as was originally envisaged  in 
\cite{Harvey:1995fq}, but only
makes the space of BPS-states into a representation of a BKM. 
\smallskip

\section{$\cN=4$ BPS Couplings and Differential Equations}
\label{GeneralAspects}
In this section we introduce and review some relevant aspects of a particular class of topological
$\mathcal{N}=4$ 
couplings $\mathcal{F}_g$ in heterotic string theory compactified on 
$\mathbb{T}^6$ (see \cite{Antoniadis:2006mr,Antoniadis:2007cw}).
In the naive field-theory limit these couplings only receive contributions 
from perturbative $1/2$ BPS-states. However, in string theory additional  
non-analytic terms appear as well. A key observation in our work is that 
one may use particular differential equations satisfied by 
$\mathcal{F}_g$ -- usually called `harmonicity equations' -- to isolate
an analytic 
part $\mathcal{F}_g^{\text{analy}}$. This represents the $\mathcal{N}=4$ analogue
of the `threshold corrections' 
in $\mathcal{N}=2$ theories. The corresponding coupling
$\mathcal{F}_g^{\text{analy}}$ will play a 
central role in the remainder of the paper. 

\setcounter{equation}{0}
\subsection{Review of $\cN=4$ Topological Amplitudes}
In \cite{Antoniadis:2006mr,Antoniadis:2007cw} (see also \cite{Antoniadis:2007ta}) a particular class of $\cN=4$
topological string 
amplitudes has been discovered. These amplitudes appear at the $g$-loop level in
type~II string 
theory compactified on $K3\times {\mathbb T}^2$, while their dual counterparts
in heterotic string theory 
compactified on ${\mathbb T}^6$ start receiving contributions at the one-loop
level. The latter expressions 
can be written as
\begin{align}
\mathcal{F}_g(y)=\int_{\mathbb{F}}\f{d^2\tau}{\bar{\eta}^{24}}\tau_2^{2g-1}
G_{g+1}(\tau,\bar{\tau})\Theta_g^{(6,22)}(\tau,\bar{\tau},y)\ ,
\label{GenTopAmplitude}
\end{align}
where the integral is over the fundamental domain $\mathbb{F}$ of
$SL(2,\mathbb{Z})$. Moreover, 
the expression
\begin{equation}
\Theta_g^{(6,22)}(\tau,\bar{\tau},y)=\left\{\begin{array}{ll}\sum_{p\in\Gamma^{6
,22}}\,
(p^L_{++})^{2g-2}\,q^{\frac{1}{2}|p^L|^2}\,\bar{q}^{\frac{1}{2}|p^R|^2} & g>1
\\[4pt]
\sum_{p\in\Gamma^{6,22}\atop  p\neq 0}\,q^{\frac{1}{2}|p^L|^2}\,\bar{q}^{\frac{1}{2}|p^R|^2} 
&g=1\end{array}\right.\label{GeneralNarainTheta}
\end{equation}
is a Siegel-Narain theta-function of the even unimodular lattice $\Gamma^{6,22}$
with momentum 
insertions $p^L_{++}$. Here the inner product on $\Gamma^{6,22}$ is defined as
in 
equation (\ref{inpro}), and 
\begin{align}
&p^L_{++}(y)=\frac{1}{2}\epsilon^{ab}\bar{u}^I_{+a}\bar{u}^J_{+b}\,p^L_{IJ}(y)\,
,&&\text{with} &&\begin{array}{l}u_I^{\pm a}\in SU(4)/S(U(2)\times U(2)) \\ y\in
SO(6,22)/SO(6)\times SO(22)\end{array}
\end{align} 
for a particular harmonic projection of the six left-moving lattice momenta (\ref{NarainMomenta}) with the 
harmonic coordinates $u_I^{\pm a}$. We will use the notation that $I=1,\ldots,4$
is an 
index of $SU(4)\cong SO(6)$, while $a=1,2$ and $\dot{a}=1,2$ are indices of
either of the 
two $SU(2)$'s, and the signs denote the charge with respect to the diagonal
$U(1)$. Finally, the 
quantities $y^{IJ}_A$ (with $A=1,\ldots, 22$ an index of $SO(22)$) span the
moduli-space
\begin{align} 
\mathcal{M}_{(6,22)}=SO(6,22)/\big(SO(6)\times SO(22)\big)
\end{align}
of the $\cN=4$ string compactification. Notice that for $g=1$ we do not sum
over 
$p=0$ in the definition (\ref{GeneralNarainTheta}). The reason for this can be
understood as 
follows. Since the amplitude $\mathcal{F}_{1}$ has no explicit $p^L$-insertions
it would 
receive contributions from $p=0$. However, integrating this contribution over
$\tau$ will 
diverge for any value of $y^{ij}_A$, thereby rendering $\mathcal{F}_{1}$
infinite. In order 
to avoid this singularity, we have chosen to regularise $\mathcal{F}_{1}$ by
performing 
the summation in (\ref{GeneralNarainTheta}) only over $p\neq 0$. This implies
that the amplitude does not receive contributions 
from massless $1/2$ BPS-states.

The object $G_{g}(\tau,\bar{\tau})$ in (\ref{GenTopAmplitude}) is a weight
$2(g+1)$ 
non-antiholomorphic modular form which can be obtained as the coefficient of 
$\lambda^{2g}$ of the generating functional \cite{Antoniadis:1995zn} (see also 
\cite{Antoniadis:2010iq})
\begin{align}
G(\lambda,\tau,\bar{\tau})=\left(\frac{2\pi i\lambda\bar{\eta}^3}
{\bar{\theta}(\lambda,\bar{\tau})}\right)^2e^{-\frac{\pi\lambda^2}{\tau_2}}\ .
\end{align}
The modular form $G_g$ can be written in terms of Eisenstein series as
\cite{Marino:1998pg}\footnote{See also 
\cite{Lerche:1987qk,Lerche:1999ju,Stieberger:1998yi,Antoniadis:2009tr} for further
examples of heterotic one-loop amplitudes involving non-holomorphic integrands.}
\begin{align}
G_g(\tau,\bar{\tau})=-\mathcal{S}_g\left(\hat{\bar{\mathcal{E}}}_2,
\frac{1}{2}
\bar{\mathcal{E}}_4,\ldots,\frac{1}{2g}\bar{\mathcal{E}}_{2g}\right)\
,\label{SchurPoly}
\end{align}
where $\mathcal{S}$ are the Schur-polynomials 
$\mc{S}_k(x_1,\dots, x_k)=x_k+\cdots + x_1^k/(k!)$ and 
\begin{align}
{\mc{E}}_{2k}(\tau)  = 2\zeta(2k) E_{2k}(\tau)
\end{align}
are the rescaled Eisenstein series of weight $2k$. Recall that
$\bar{\mathcal{E}}_{2}$ is a 'quasi-modular form' \cite{Dijkgraaf,KanekoZagier}, implying that it does not 
only transform with a weight under modular transformations but receives an
additional 
anomalous shift-term. Following standard practice we have therefore introduced the quantity 
\begin{align} 
\mc{\hat{\bar{E}}}_2(\tau, \bar\tau)= \f{\pi^2}{3}
\left(E_2(\bar{\tau})-\f{3}{\pi \tau_2}\right)\ ,
\end{align}
which is an honest weight 2 modular form, but is non-antiholomorphic in $\tau$. 

In \cite{Antoniadis:2007cw} the amplitudes (\ref{GenTopAmplitude}) were shown to
compute BPS 
couplings in the string 
effective action, which in harmonic superspace take the form (see
\cite{Antoniadis:2007cw} for further details)
\begin{align}
S=\int d^4x\int du\int d^4\theta^+\int
d^4\bar{\theta}_-(K^{++}_{\mu\nu}K^{++,\mu\nu})^{g+1}\mathcal{F}_{g}(Y^{++}_A,
u)\ .\label{BPScoupling}
\end{align}
Here $K^{IJ}_{\mu\nu}$ is a particular super-descendant of the (linearised)
$\cN=4$ supergravity 
multiplet, while $Y^{IJ}_{A}$ is a linearised $\cN=4$ vector-multiplet, whose
lowest components 
$y^{IJ}_A$ form the moduli space $\mathcal{M}_{(6,22)}$ of the $\cN=4$ string
compactification. 

\subsection{Differential Equations for $g>1$}
\label{Section:HarmonicityGenusg}
As was shown in \cite{Antoniadis:2007cw}, for $g>1$ the amplitudes
(\ref{GenTopAmplitude}) satisfy 
certain differential equations with respect to the moduli of the heterotic
$\cN=4$ compactification. In 
particular
\begin{align}
&\epsilon_{ab}\epsilon^{IJKL}\frac{\partial}{\partial\bar{u}^J_{+b}}D_{KL,A}
\mathcal{F}_g=(2g-2)\bar{u}^I_{+a}D_{++,A}\mathcal{F}_{g-1}
\label{HarmonicityRelation}\\
&\left(\epsilon^{IJKL}D_{IJ,A}D_{KL,B}+4(g+1)\delta_{AB}\right)
\mathcal{F}_g=4D_{++,A}D_{++,B}\mathcal{F}_{g-1}\ ,\label{SecondOrderRelation}
\end{align}
where $D_{++,A}$ are harmonic projections of the covariant derivatives
$D_{ij,A}$ in the moduli space 
$\mathcal{M}_{(6,22)}$. We will refer to these equations as the harmonicity and
second order relation, 
respectively. 

Note that in both equations the amplitude $\mathcal{F}_{g-1}$ appears on the
right hand side.
As for the holomorphic anomaly equation (see {\it e.g.}\
\cite{Bershadsky:1993cx}), we shall
call these contributions anomalous. From the string effective action point of
view they can be 
understood as arising via the violation of certain analyticity properties of the
corresponding 
BPS-couplings. To understand this, we recall that the coupling
(\ref{BPScoupling}) is half-BPS 
in the sense that the integrand is annihilated by half of the spinor-derivatives
of the $\cN=4$ 
harmonic superspace (`G-analyticity constraint'). For this to be true, however,
it is essential that 
$\mathcal{F}_g$ is a function of only a particular projection of the vector
multiplets, namely 
$Y^{++}_A=\epsilon_{ab} u_I^{+a}u_J^{+b}Y^{IJ}_A$. As was explained in
\cite{Antoniadis:2007cw}, 
this particular dependence leads to the differential equations
(\ref{HarmonicityRelation}) and 
(\ref{SecondOrderRelation}), however, with the right hand side replaced by
zero. 
The appearance of $\mathcal{F}_{g-1}$ in the explicit string computation can
therefore 
be understood as an anomalous violation of these analyticity constraints.

To understand how the anomalous terms in (\ref{HarmonicityRelation}) and 
(\ref{SecondOrderRelation}) arise from the string amplitude
(\ref{GenTopAmplitude}) we 
recall from \cite{Antoniadis:2007cw} 
\begin{align}
&\epsilon_{ab}\epsilon^{IJKL}\frac{\partial}{\partial\bar{u}^J_{+b}}D_{KL,A}
\mathcal{F}_g
\nonumber\\
&\hspace{0.5cm}=4i(2g-2)\bar{u}^I_{+a}\int \f{d^2\tau}{\bar{\eta}^{24}}
G_{g+1}(\tau,\bar{\tau})\frac{\partial}{\partial\tau}\left[\tau_2^{2g}
\sum_{p\in \Gamma^{6,22}}(p^L_{++})^{2g-3}p^R_Aq^{\frac{1}{2}|p^L|^2}
\bar{q}^{\frac{1}{2}|p^R|^2}\right]\ ,
\end{align}
and the second order equation
\begin{align}
&\left(\epsilon^{IJKL}D_{IJ,A}D_{KL,B}+4(g+1)\delta_{AB}\right)\mathcal{F}_g\nonumber\\
&\hspace{0.5cm}=-32\pi i\int \frac{d^2\tau}{\bar{\eta}^{24}}
G_{g+1}(\tau,\bar{\tau})\frac{\partial}{\partial\tau}\left[\tau_2^{2g+1}
\sum_{p\in\Gamma^{6,22}}\left(p^R_Ap^R_B
-\frac{\delta_{AB}}{4\pi\tau_2}\right)(p^L_{++})^{2g-2}q^{\frac{1}{2}|p^L|^2}
\bar{q}^{\frac{1}{2}|p^R|^2}\right]\ .\label{SecOrderpart}
\end{align}
Notice that in both cases, since $g>1$, after performing a partial integration,
the boundary term 
vanishes and the anomaly stems from the contribution which is proportional to
\begin{align}
\frac{\partial}{\partial\tau}
G_{g+1}(\tau,\bar{\tau})=-\frac{i\pi}{2\tau_2^2}G_{g}(\tau,\bar{\tau})\ .
\end{align}
Recalling the expression (\ref{SchurPoly}) for $G_{g}$ in terms of Schur
polynomials, we deduce that the 
only source of non-antiholomorphicity is the explicit dependence on $\tau_2$ in 
$\hat{\bar{\mathcal{E}}}_2$. Therefore, we can split 
\begin{align}
G_g(\tau,\bar{\tau})=G_g^{\text{analy}}(\bar{\tau})+G_g^{\text{non-analy}}(\tau,
\bar{\tau}) \label{BPSgfunct}
\end{align}
with the explicit expressions
\begin{align}
&G_g^{\text{analy}}(\bar{\tau})=-\mathcal{S}_g\left(0,\frac{1}{2}\bar{\mathcal{E
}}_4,\ldots,\frac{1}{2g}\bar{\mathcal{E}}_{2g}\right)\  , \\
&G_g^{\text{non-analy}}(\tau,\bar{\tau})=-\mathcal{S}_g\left(\hbe,\frac{1}{2}
\bar{\mathcal{E}}_4,\ldots,\frac{1}{2g}\bar{\mathcal{E}}_{2g}\right)
+\mathcal{S}_g\left(0,\frac{1}{2}\bar{\mathcal{E}}_4,\ldots,\frac{1}{2g}\bar{
\mathcal{E}}_{2g}\right)\ ,
\end{align}
which both have weight $2g$ under modular transformations. The anomaly of 
(\ref{HarmonicityRelation}) and (\ref{SecondOrderRelation}) (and therefore also
the violation of 
G-analyticity) can now be fully (and uniquely) attributed to
$G_g^{\text{non-analy}}$. 
It is therefore consistent to define the purely analytic contribution to the
amplitude as 
\begin{align}
\mathcal{F}_g^{\text{analy}}(y)=\int_{\mathbb{F}}\f{d^2\tau}{\bar{\eta}^{24}}
\tau_2^{2g-1}
G_{g+1}^{\text{analy}}(\bar{\tau})\Theta_g^{(6,22)}(\tau,\bar{\tau},y)\
,\label{BPSamplitudeDef}
\end{align}
which yields a vanishing anomaly when inserted into the harmonicity and second
order relation.
\subsection{Differential Equations for $g=1$}
\label{Section:HarmonicityGenus1}
In this paper we will mostly be interested in the amplitude
(\ref{GenTopAmplitude}) for $g=1$; the case $g=1$ is somewhat more subtle and 
requires special care. First of all, the 
right hand side of the harmonicity relation (\ref{HarmonicityRelation})
vanishes, which reflects the 
fact that $\mathcal{F}_{1}$ is independent of the harmonic variables
$\bar{u}^i_{\pm a}$, as can 
also be seen from the effective action coupling (\ref{BPScoupling}). Focusing on
the remaining 
second order relation (\ref{SecondOrderRelation}), we notice that for $g=1$ a
partial integration 
of (\ref{SecOrderpart}) will also produce a non-trivial boundary contribution at
$\tau_2\to\infty$ 
for those points of the lattice for which $p^L=0$. Explicitly we find
\begin{align}
&\left(\epsilon^{IJKL}D_{IJ,A}D_{KL,B}+8\delta_{AB}\right)\mathcal{F}_1\nonumber\\
&\hspace{1cm}=-32\pi i\lim_{\tau_2\to\infty}\int_{-\frac{1}{2}}^{\frac{1}{2}} 
\f{d\tau_1}{\bar{\eta}^{24}}\tau_2^3 G_2(\tau, \btau)
\sum_{ p\in\Gamma^{6,22}\atop p\neq 0}\left(p^R_Ap^R_B
-\frac{\delta_{AB}}{4\pi\tau_2}\right)q^{\frac{1}{2}|p^L|^2}\bar{q}^{\frac{1}{2}
|p^R|^2}\nonumber\\
&\hspace{1.5cm}+16\pi^2\int \f{d^2\tau}{\bar{\eta}^{24}}\tau_2^3
\left(\frac{\partial}{\partial\tau}G_{2}^{\text{non-analy}}(\tau,\bar{\tau}
)\right)
\sum_{p\in\Gamma^{6,22}\atop p\neq 0}\left(p^R_Ap^R_B
-\frac{\delta_{AB}}{4\pi\tau_2}\right)q^{\frac{1}{2}|p^L|^2}\bar{q}^{\frac{1}{2}
|p^R|^2}\ .
\label{Genus1Anomaly}
\end{align}
The last line arises by the same mechanism as just discussed for the case $g>1$,
and it will 
vanish if we restrict $\mathcal{F}_1$ to its analytic part
$\mathcal{F}_1^{\text{analy}}$. 
The first line, however, is an additional contribution for which we can write
\begin{align}
&\left(\epsilon^{IJKL}D_{IJ,A}D_{KL,B}+8\delta_{AB}\right)\mathcal{F}_1^{\text{
analy}}
=8i\delta_{AB}\lim_{\tau_2\to \infty}\int_{-\frac{1}{2}}^{\frac{1}{2}}
\frac{d\tau_1}{\bar{\eta}^{24}}\tau_2^2 G_2^{\text{analy}}(\bar{\tau})
\sum_{p\in\Gamma^{6,22}\atop p\neq 0}\bar{q}^De^{-\pi\tau_2|p^L|^2}\,, \label{Divergence}
\end{align}
where we have extended the definitions (\ref{BPSgfunct}) and
(\ref{BPSamplitudeDef}) 
to the case $g=1$, and $D$ was defined in (\ref{discrim}). At a generic point in moduli space ({\it i.e.}\
for generic $y^{IJ}_A$) 
the limit will simply vanish.\footnote{As we will see in the following section~\ref{SingularityStructure}, at very particular points in the moduli space, (\ref{Divergence}) may diverge for $p_L=0$ and $D=1$. At these particular points the differential equation will only be consistent after
some additional proper regularisation. For determining the analytic part of the integral, however, we will assume to work at a generic point in the moduli space.} Therefore we can (as in the case of $g>1$) attribute the
anomaly completely to the term
\begin{align}
&G_{2}^{\text{non-analy}}(\tau,\bar{\tau})=-\mathcal{S}_2\left(\hbe,\frac{1}{2}
\bar{\mathcal{E}}_4\right)+\mathcal{S}_2\left(0,\frac{1}{2}\bar{\mathcal{E}}
_4\right)\ .
\end{align}


\section{Singularities of the Analytic Integral}\label{SingularityStructure}
\setcounter{equation}{0}

In this section we shall study explicitly the analytic part of the genus one
topological amplitude 
$\mathcal{F}_{g=1}$ discussed in the previous section. We shall analyse
in detail 
the structure of the singularities of the integral as a function of the moduli 
$y=(U,T;\vec{V})\in \mathbb{C}^{1,17}$. This will reveal the first piece of evidence for a 
relation to the  BKM algebra $\bor$ introduced in section~\ref{AlgebraBPSstates}.
In particular,  we shall show that at least certain singularities are associated with Weyl
reflections of the Lorentzian Kac-Moody algebra $({\mathfrak e}_8\oplus {\mathfrak e}_8)^{++}$ 
whose Borcherds lift is $\bor$.

\subsection{General Analysis of the Singularities in Eight Dimensions}
\label{largevol}

In order to make contact with 
the discussion in section~\ref{AlgebraBPSstates} we will consider the 
internal manifold to be factorised as ${\mathbb T}^6={\mathbb T}^4\times
{\mathbb T}^2$, 
and take the large volume limit of ${\mathbb T}^4$. On the level of the integral
these assumptions 
mean that the Siegel-Narain theta function 
of the original $\Gamma^{6,22}$ Narain-lattice will be decomposed as
\begin{align}
\frac{G_{2}(\tau,\bar{\tau})\tau_2^2}{\bar{\eta}^{24}}\,\Theta^{(6,22)}_{g=1}
\sim\text{Vol}\,\frac{G_{2}(\tau,\bar{\tau})}{\bar{\eta}^{24}}\,\Theta^{(2,18)}_{g=1}(\tau,\bar{\tau},y) \ ,
\label{PreReductionSiegelNarain}
\end{align}
where $\text{Vol}$ is the volume of ${\mathbb T}^4$ and 
$\Theta^{(2,18)}_{g=1}(\tau,\bar{\tau},y)$ is the corresponding 
Siegel-Narain theta function for the lattice $\Gamma^{2,18}$.  Furthermore, the
contribution from the analytic part of $G_{2}(\tau,\bar{\tau})$ can be rewritten as
\begin{equation}
\frac{G_2^{\rm analy}(\tau,\bar{\tau})}{\eta^{24}(\bar{\tau})} =
-\zeta(4) \frac{E_4(\bar{\tau})}{\eta(\bar{\tau})^{24}}\equiv -\zeta(4)\mathcal{P}(\bar{\tau})\ . 
\label{defP}
\end{equation}
We shall drop the irrelevant overall factor of $-\zeta(4)$ from now on and write the integral 
$\mathcal{F}_1^{\text{analy}}$ as 
\begin{align}
&\mathcal{F}_{1}^{\text{analy}}
=\int_{\mathbb{F}}\frac{d^2\tau}{\tau_2}\,
\mathcal{P}(\bar{\tau})
\sum_{p\in\Gamma^{2,18}\atop p\neq 0}\bar{q}^De^{-\pi\tau_2|p^L|^2}\ ,
\label{1loopBPSIntro}
\end{align}
where $D=\frac{1}{2}\left(|p^R|^2-|p^L|^2\right)$ was introduced in
(\ref{discrim}), and 
$p=(m_1,n_1;m_2,n_2;\vec{\ell}\,\,)$ with 
$\vec{\ell}=(\vec{\ell}_1,\vec{\ell}_2)\in\Lambda_{\mathfrak{e}_8}\oplus \Lambda_{\mathfrak{e}_8}$. 
To understand the mechanism by which a singularity might occur in
(\ref{1loopBPSIntro}) 
(see also
\cite{Borcherds1,Harvey:1995fq,Borcherds2,LopesCardoso:1996nc,Obers:1999um,%
David:2006ud,Cheng:2008fc} for related discussions), 
we observe that in the standard fundamental domain $\mathbb{F}$ of
$SL(2,\mathbb{Z})$, the upper boundary of the $\tau_2$ integration is at infinity. At a generic point in
moduli space, the integrand is exponentially suppressed as $\tau_2\to\infty$ due to the
exponentials of the Narain momenta. However, at particular points in moduli space this damping might fail,
thus leaving an unregulated integral which ultimately leads to a logarithmic divergence. A
necessary condition for such a divergence to appear is
\begin{align}
|p^L|=0\ ,\label{SingularityConstraint}
\end{align}
since then the suppression induced by the factor $e^{-\pi\tau_2|p^{L}|^2}$ in
(\ref{1loopBPSIntro}) is absent. However, condition (\ref{SingularityConstraint}) 
alone is not sufficient since the integrand also involves a power series in $\bar{q}^{\; m}$, and 
for $|p^L|=0$, the $\tau_1$-integral picks out the constant term, $\bar{q}^{\; 0}$. Fourier 
expanding the integrand
\begin{align}
\mathcal{P}(\bar{\tau}) 
\sum_{p\in\Gamma^{2,18}\atop p\neq 0 ,\, p^{L}=0}
\bar{q}^{\frac{1}{2} |p^R|^2} =\sum_{p\in\Gamma^{2,18}\atop p\neq 0 ,\, 
p^{L}=0}\sum_{n=-1}^\infty d(n)\bar{q}^{n+D}
\end{align}
we therefore only encounter a divergence if $d(-D)\neq 0$ for some vector in the 
sum over $\Gamma^{2,18}$. In the following we shall focus on 
those terms for which $D=1$, {\it i.e.}\ the terms that arise from the $\bar{q}^{-1}$ term of  
$1/\eta(\bar{\tau})^{24}$ in (\ref{defP}); these singularities
will be directly related to the Weyl reflections in $\bor$. Given the construction of section~2,
the other singularities (with $D<1$) should then have an interpretation in terms 
of  $\boru$ --- this takes into account the overall 
$E_4(\bar{\tau})=\Theta_{\mathfrak{e}_8}(\bar\tau)$ factor in (\ref{defP}).

In order to study this problem, let us 
fix a $p\in\Gamma^{2,18}$ with $\langle p|p\rangle=2$ (so that $D=1$), and ask for which  
values of the moduli $y=(U,T;\vec{V})$ its contribution to the sum leads to a divergence, 
{\it i.e.}\ for which $y$ we have $p^L=0$. Because of (\ref{discrim}), we have the  equivalence
\begin{equation}\label{pLcond}
p^L = 0 \quad \Longleftrightarrow \quad \langle p | u(y) \rangle = 0 \ . 
\end{equation}
The actual moduli space is parametrised by $y$, where we identify $y\sim y'$ if 
$u(y) = A u(y')$ for $A\in SO(2,18;{\mathbb Z})$; this is the familiar T-duality
action. Next we observe 
that the inner product (\ref{pLcond}) is invariant under this T-duality action, 
{\it i.e.} 
\begin{equation}
\langle A p | A u(y) \rangle = \langle p | u(y) \rangle \quad \text{for all $\
A\in SO(2,18;{\mathbb Z})$.}
\end{equation}
Thus if (\ref{pLcond})  is satisfied for $p=p_1$ at $y=y_1$, and  we consider
$p_2 = A p_1$ with 
$A\in SO(2,18;{\mathbb Z})$, then (\ref{pLcond})  vanishes for $p=p_2$ at
$y=y_2\sim y_1$
since $u(y_2) = A u(y_1)$. It is therefore sufficient to consider one
representative of $p$ for 
each $SO(2,18;{\mathbb Z})$-orbit.

It was shown in \cite{Wall} (see also \cite{Banerjee:2007sr}) that for every
$p\in \Gamma^{2,18}$
with $\left<p |p\right> = 2$, there exists an $SO(2,18;{\mathbb Z})$
transformation that maps it to 
$\hat{p} = A p\in\Gamma^{1,17}$, {\it i.e.}\
\begin{equation}\label{4.8}
\hat{p} = (m_1,n_1;m_2=0,n_2=0;\vec{\ell}\,\,) \ .
\end{equation}
It is therefore sufficient to restrict ourselves to such vectors.  For those the
analysis of $\hat{p}^L=0$ 
is  now straightforward since (\ref{discrim}) implies that $\hat{p}^L=0$ is
equivalent to 
\begin{equation}\label{eq1}
m_1 T + n_1 U - \vec{\ell}\cdot \vec{V} = 0 \ ,
\end{equation}
while the constraint $\langle p| p \rangle =  2$ leads  to 
\begin{equation}\label{eq2}
\vec{\ell}^{\ 2} - 2 m_1 n_1=\vec{\ell}_1^{\ 2}+\vec{\ell}_2^{\ 2} - 2 m_1 n_1 = 2 \ .
\end{equation}
This condition now has a nice Lie algebraic interpretation: since
$\hat{p}\in\Gamma^{1,17}= \Pi^{1,1}\oplus \Lambda_{\mathfrak{e}_8} \oplus \Lambda_{\mathfrak{e}_8}$, 
we can think of  $\hat{p}$ as an element of the root lattice of the double
extension $\mathfrak{g}^{++}$ of 
$\mathfrak{g} = {\mathfrak e}_8\oplus {\mathfrak e}_8$. 
The constraint $\left<\hat{p}| \hat{p}\right>=(\alpha|\alpha)=2$ implies that 
$\alpha=\hat{p}$ is a real root of $\mathfrak{g}^{++}$, and the condition 
(\ref{pLcond}) is then equivalent to the statement that $u(y)$ is a fixed
point of the Weyl reflection $w_\alpha \in \mathcal{W}({\mathfrak g}^{++})$ 
with respect to the root $\alpha$. Here $w_\alpha$ acts on the moduli vector 
$y=(U,T;\vec{V})\in \mathbb{C}^{1,17}$ as 
\be
w_\alpha \ :\ y \ \longmapsto\  y - (y|\alpha) \, \alpha \ ,
\label{weyl}
\ee
where $(\cdot |\cdot )$ is the inner product on the lattice 
$\Lambda_{\mathfrak{g}_{++}}$, 
\be
(\alpha| \alpha)=-2m_1n_1+\vec{\ell}\cdot \vec{\ell}=-2m_1n_1+\vec{\ell}_1^{\ 2}+\vec{\ell}_2^{\ 2}\ ,
\ee
which is the one inherited from $\left< \cdot |\cdot \right> $ on
$\Gamma^{2,18}$. 
Note that for vectors of the type $\hat{p}$ (for which $m_2=n_2=0$) 
it does not matter whether we take the inner product with $y$ or $u(y)$.
Thus we conclude that singularities of the BPS integral occur precisely at
the fixed points of the Weyl group $\mathcal{W}({\mathfrak g}^{++})$.

\subsection{Explicit Singular Loci}\label{Sect:SingMassless}

Let us describe the relevant Weyl group $\mathcal{W}({\mathfrak g}^{++})$ 
and its singularities more explicitly. The Lie algebra $\mathfrak{g}^{++}$ is 
the double extension of $\mathfrak{g} = \mathfrak{e}_8\oplus\mathfrak{e}_8$. 
As explained in appendix~\ref{App:ExtendSemiSimple}, the construction of
$\mathfrak{g}^{++}$ naturally involves an `auxiliary' algebra
${\tilde{\mathfrak{g}}^{++}}$, from which the physically relevant
double extension $\mathfrak{g}^{++}$ is obtained by taking the quotient
${\tilde{\mathfrak{g}}^{++}}/\mathfrak{r}$ by the center $\mathfrak{r}$.
Let us begin by introducing a basis of simple roots $\tilde{\alpha}_I$ for
${\tilde{\mathfrak{g}}^{++}}$
\begin{equation}\label{SetRootsOver1}
\begin{array}{lll}
{\tilde{\alpha}}_{-1}=(1,-1;\vec{0};\vec{0}) \qquad
&{\tilde{\alpha}}_0^{(1)}=(-1,0;-\vec{\theta};\vec{0}) \qquad 
&{\tilde{\alpha}}_0^{(2)}=(-1,0;\vec{0};-\vec{\theta}) \\ 
{\tilde{\alpha}}_i^{(1)}=(0,0;\vec{e}_i;\vec{0}) \qquad
&{\tilde{\alpha}}_i^{(2)}=(0,0;\vec{0};\vec{e}_i) \ , & 
\end{array}
\end{equation}
where $\vec{e}_i$ is a basis of simple roots ($i=1,\ldots, 8$) for ${\mathfrak e}_8$, and
$\vec{\theta}$ the corresponding highest root.
The roots (\ref{SetRootsOver1})  define an overcomplete basis 
for the root lattice
$\Lambda_{{\mathfrak{g}^{++}}}=\Pi^{1,1}\oplus\Lambda_{\mathfrak{e}_8}\oplus
\Lambda_{\mathfrak{e}_8}$. In fact, there is one relation (generating the center $\mathfrak{r}$ 
of ${\tilde{\mathfrak{g}}^{++}}$,  see appendix~\ref{App:ExtendSemiSimple}) which we may 
use to express $\tilde{\alpha}^{(2)}_0$ in terms of the other roots 
\begin{align}
&\tilde{\alpha}^{(2)}_0=\tilde{\alpha}^{(1)}_0
+\sum_{i=1}^{8}(\vec{\theta}\cdot \vec{f}^{\; i})\, (\tilde{\alpha}^{(1)}_i-\tilde{\alpha}^{(2)}_i)\ .
\label{SemiSimpExpressRoot}
\end{align}
Here $\vec{f}^{\; i}$ are the fundamental weights of
$\mathfrak{e}_{8}$. Hence, we can then write for any $\alpha\in\Lambda_{\mathfrak{g}^{++}}$
\begin{align}
\alpha=x_{-1}{\tilde{\alpha}}_{-1}+x_0\tilde{\alpha}_0^{(1)}+\sum_{i=1}^{8}\left(x_i^{
(1)}{\tilde{\alpha}}_i^{(1)}+x_i^{(2)}{\tilde{\alpha}}_i^{(2)}\right)\,,
\label{GenElRootSem}
\end{align}
with integer coefficients $(x_{-1},x_0;\vec{x}^{\,(1)},\vec{x}^{\,(2)})$. Using the 
same inner product as in (\ref{innerProdRed}) we find that the product
between $\alpha$ and a moduli vector $y=(U,T;\vec{V}_{(1)}, \vec{V}_{(2)})$ reads
\begin{align}
({\alpha}|y)=x_{-1}T+(x_0-x_{-1})U&-x_0\,(\vec{\theta}\cdot\vec{V}_1)+
\sum_{i=1}^{8}\left[x_i^{(1)}(\vec{e}_i\cdot\vec{V}_{(1)})+x_i^{(2)}(\vec{e}_i\cdot\vec{V}_{(2)})\right]
\ .\label{semisimpleinnerproduct}
\end{align}
Thus there are $18$ linearly independent singular divisors 
\begin{equation} \label{semisimplewalls} 
\begin{array}{rcl}
\tilde{{\cal D}}_{-1} & = & \left\{ y \in {\mathcal M}_{2,18} \ | \
(y|\tilde{\alpha}_{-1})=U-T=0 \right\}   \\[2pt]
\tilde{{\cal D}}_{0}\phantom{.} & = & \left\{ y \in {\mathcal M}_{2,18} \ | \ (y|\tilde{\alpha}^{(1)}_0)
= T - \vec{\theta} \cdot \vec{V}_{(1)}= 0 \right\} \\
\tilde{{\cal D}}^{(a)}_{i} & =& \left\{ y \in {\mathcal M}_{2,18} \ | \ 
(y|\tilde{\alpha}^{(a)}_i)=\vec{e}_i \cdot \vec{V}_{(a)}= 0 \right\} \qquad 
\begin{array}{l}a=1,2  \\ i=1,\ldots,8\ , \end{array} 
\end{array}
\end{equation}
which we will sometimes collectively denote by $\tilde{\mathcal{D}}_I$ with $I=-1,\ldots,16$. 
The divisor $\tilde{\cal D}_{-1}$ is independent of the Wilson line, and thus the 
corresponding singularity of the integral cannot be removed by a shift of
$\vec{V}$. In fact, this is exactly the locus of enhanced gauge symmetry which 
was for example discussed in \cite{LopesCardoso:1994ik,Harvey:1995fq,Borcherds2}.

For a given point $y$ in the moduli space  $\mathbb{C}^{1,17}$ the 
divisors $\tilde{\cal D}_I$ represent the `dominant' walls of the complexified Weyl
chamber, in the sense that all other walls lie `behind' this set of walls.  If
we restrict the moduli to the fundamental Weyl chamber
\be
\mathcal{C}_{\mathbb{C}} = \{ y\in   \mathbb{C}^{1,17}\ |\ \tilde{\cal D}_I\geq 0 \} \ , 
\label{cplxWeylchamber}
\ee
the only singularities appear at the boundary of $\mathcal{C}_{\mathbb{C}}$. 
Note that since we are working with the complexified Weyl chamber, eq.\
(\ref{cplxWeylchamber}) should be understood as providing separate conditions on the real 
and imaginary parts of the moduli vector $y=(U,T;\vec{V})$.


\section{BPS Amplitude and Denominator Identity}\label{Section:Akintegral}
\setcounter{equation}{0}

Our next aim is to evaluate the 1-loop integral (\ref{1loopBPSIntro}) explicitly. As we shall see,
the analytic part $\mathcal{F}_{g=1}^{\text{analy}}$ can be related to the infinite product side of the 
denominator formula for the Borcherds algebra $\boru$ which contains the BPS symmetry
algebra $\bor$ as a subalgebra.

\subsection{Torus Integral}\label{Sect:TorusIntegralEval}
We now use the methods developed in \cite{Dixon:1990pc} and 
further extended in \cite{Harvey:1995fq} (see also 
\cite{LopesCardoso:1996nc,Foerger:1998kw,Lerche:1998nx,Obers:1999um}) 
to tackle the $\tau$ world-sheet torus integral. The moduli dependence is described
by the  Siegel-Narain theta function of the lattice $\Gamma^{2,18}$
\begin{align}
&\Theta^{(2,18)}(\tau,\bar{\tau};y)
=\sum_{ x\neq(0,0;0,0;\vec{0})}\bar{q}^{\frac{1}{2}\,\langle x|x\rangle}\,
e^{2\pi \tau_2\frac{|\langle x|u(y)\rangle|^2}{(\Im y|\Im
y)^2}}\  ,\label{SiegelNarain22k}
\end{align}
for which we shall use the same notation as in section~\ref{AlgebraBPSstates} and 
parametrise the summation by $x = (m_1,n_1;m_2,n_2;\vec{\ell}\,\,)$ with 
$\vec{\ell}\in \Lambda_{\mathfrak{e}_8}\oplus\Lambda_{\mathfrak{e}_8}$. We can perform 
a Poisson resummation on the indices $m_{1,2}$
\begin{align}
\Theta^{(2,18)}=&\sum_{{(p_1,
n_1;p_2,n_2)}\atop{\vec{\ell}\in \Lambda_{\mathfrak{e}_8}\oplus\Lambda_{\mathfrak{e}_8}}}
\int_{-\infty}^{\infty}\!du_{1,2}\,\bar{q}^{\frac{1}{2}\vec{\ell}\cdot
\vec{\ell}-u_1n_1-u_2n_2}\, 
e^{2\pi i(p_1u_1+p_2u_2)-\frac{\pi\tau_2}
{(\Im y|\Im y)}\left|u_2+u_1U+n_1T-\frac{n_2}{2}\,(y|y)+\vec{\ell}\cdot
\vec{V}\right|^2}\ .\nonumber
\end{align}
Both $u$-integrals are of Gaussian type and can therefore be performed
using elementary 
methods. Thus we get  (see also
\cite{Dixon:1990pc,Harvey:1995fq,LopesCardoso:1996nc,Foerger:1998kw,Lerche:1998nx,Obers:1999um})
\begin{align}
\mathcal{F}_{g=1}^{\text{analy}}=\int_{\mathbb{F}}\frac{d^2\tau}{\tau_2^2}
\frac{Y}{U_2}\,\mathcal{P}(\bar{\tau})\!\!
\sum_{{(p_1,n_1;p_2,n_2)}\atop\vec{\ell}\in \Lambda_{\mathfrak{e}_8}\oplus\Lambda_{\mathfrak{e}_8}}\,
&\bar{q}^{\frac{1}{2}\vec{\ell}\cdot \vec{\ell}}\,e^{2\pi i\vec{\ell}\cdot
\vec{z}-\frac{\pi Y}
{U_2^2\tau_2}|\mathcal{A}|^2-2\pi iT\text{det}A-\frac{\pi n_2
\left(\vec{V}^2\tilde{\mathcal{A}}-\vec{\bar{V}}^2\mathcal{A}\right)}{U_2}
+\frac{2\pi i\,(\Im\vec{V})^2}{U_2^2}(n_1+n_2\bar{U})\mathcal{A}}, \label{IntegralbeforeOrbits}
\end{align}
where $\mathcal{P}(\bar{\tau})$ was defined in (\ref{defP}),
\begin{align}
&A=\left(\begin{array}{cc}n_1&-p_1 \\ n_2 & p_2\end{array}\right)\ ,
&&\mathcal{A}=\left(1,U\right)A\left(\begin{array}{c}-\bar{\tau} \\
1\end{array}\right) \ ,
&& \tilde{\mathcal{A}}=\left(1,\bar{U}\right)A\left(\begin{array}{c}-\bar{\tau}
\\ 1\end{array}\right)\ ,
\end{align}
and 
\begin{equation}
Y=(\Im y|\Im y) \qquad  \qquad
\vec{z}=\frac{i}{2 U_2}\,(\vec{V}\tilde{\mathcal{A}}-\vec{\bar{V}}\mathcal{A})\ . 
\end{equation}
For computing the $\tau$-integration it is convenient to introduce the Fourier expansion 
\begin{align}
\mathcal{P}(\bar{\tau})
\sum_{\vec{\ell}\in\Lambda_{\mathfrak{e}_8}\oplus\Lambda_{\mathfrak{e}_8}}\,
\bar{q}^{\frac{1}{2}\vec{\ell}\cdot \vec{\ell}}\,
e^{2\pi i\vec{\ell}\cdot \vec{z}}=\sum_{n=-1}^\infty
\sum_{\vec{\ell}\in\Lambda_{\mathfrak{e}_8}\oplus\Lambda_{\mathfrak{e}_8}} 
\bar{c}_{\rm ext}(n-\tfrac{1}{2}\vec{\ell}^{\;\, 2}\;)\,\bar{q}^n\,e^{2\pi i \vec{\ell}\cdot
\vec{z}}\ ,\label{FourierExpansionConjugacy}
\end{align}
where we have used again that the left hand side is a weak Jacobi form and thus the Fourier coefficients
only depend on $(n,\vec{\ell}\,\,)$ through the combination $n-\tfrac{1}{2}\vec{\ell}^{\;\, 2}$.
Note that the coefficients agree precisely with those appearing in the Fourier expansion 
(\ref{GeneralCountingFormula}) after restricting one of the theta series to its associated `theta constant', 
{\it i.e.}\ 
\be\label{rest}
 \sum_{n=-1}^\infty
\sum_{\vec{\ell}\in\Lambda_{\mathfrak{e}_8}\oplus\Lambda_{\mathfrak{e}_8}} 
\bar{c}_{\rm ext}(n-\tfrac{1}{2}\vec{\ell}^{\; 2}\,)\,\bar{q}^n\,e^{2\pi i \vec{\ell}\cdot
\vec{z}}=\frac{\Theta_{\mathfrak{e}_8\oplus\mathfrak{e}_8}(\bar{q}, \vec{z}) 
\Theta_{\mathfrak{e}_8}(\bar{q}, \vec{0})}{\eta(\bar{q})^{24}}\ ,
\ee
where $\Theta_{\mathfrak{e}_8}(\bar{q}, \vec{0})=E_4(\bar{\tau})$. The 
coefficients $\bar{c}_{\rm ext}$ can therefore be expressed in terms of  the
$c_{\text{ext}}$ as 
\be
\bar{c}_{\text{ext}}(n-\tfrac{1}{2}\vec{\ell}^{\, 2}\,)= \sum_{\vec{\alpha}\in\Lambda_{\mathfrak{e}_8}}
c_{\text{ext}}(n,\vec{\ell},\vec{\alpha}) \ ,
\label{cextbar}
\ee
and thus can be interpreted as `averages' over the root lattice of ${\mathfrak e}_8$. 
The first few terms are explicitly
\begin{align}
&\bar{c}_{\rm ext}(-1)=1\ ,
&&\bar{c}_{\rm ext}(0)=264\ ,
&&\bar{c}_{\rm ext}(1)=8244\ ,
&&\bar{c}_{\rm ext}(2)=139520\ ,
\end{align}
and $\bar{c}_{\rm ext}(n)=0$ for $n<-1$.

With these preparations out of the way we can now compute the $\tau$-integral in 
(\ref{IntegralbeforeOrbits}) following 
\cite{Dixon:1990pc,Harvey:1995fq,Kawai:1995hy,LopesCardoso:1996nc,Stieberger:1998yi,Foerger:1998kw}. 
Using modular invariance of the integrand, we can trade a modular transformation
$\tau\mapsto\frac{a\tau+b}{c\tau+d}$ for a transformation of the matrix 
\begin{equation}
A\mapsto A\left(\begin{array}{cc} a & b \\ c & d\end{array}\right) \ .
\end{equation}
This allows us to extend the domain of integration to images of 
$\mathbb{F}$ under $SL(2,\mathbb{Z})$, while simultaneously restricting the
summation over $A$
to inequivalent $SL(2,\mathbb{Z})$-orbits. As was first 
discussed in \cite{Dixon:1990pc}, there are three inequivalent contributions
\begin{align}
\mathcal{F}_{g=1}^{\text{analy}}=\mathcal{I}^{(\text{analy})}_{0}+\mathcal{I}_{
ND}^{(\text{analy})}+\mathcal{I}_{D}^{(\text{analy})}\
,\label{IntegralSummaryRes}
\end{align}
corresponding to three different classes of representatives of the matrix $A$,
called the zero, 
non-degenerate and degenerate orbits, respectively. The computation of each of
these separately 
is rather tedious but follows quite closely
\cite{Dixon:1990pc,Harvey:1995fq,LopesCardoso:1996nc,Foerger:1998kw}. 
We have relegated these calculations to appendix~\ref{App:Orbits}, and the final
result is 
\begin{align}
\mathcal{F}_{g=1}^{\text{analy}}
=&\sum_{\vec{\ell}\in\Lambda_{\mathfrak{e}_8}\oplus\Lambda_{\mathfrak{e}_8}}
\bigg[\frac{2\pi
Y}{3U_2}\left(\bar{c}_{\rm ext}(0,\vec{\ell}\,\,)
-24\bar{c}_{\rm ext}(-1,\vec{\ell}\,\,)\right)
+2\log\left|1-e^{2\pi i\vec{\ell}\odot\vec{V}}\right|^{\bar{c}_{\rm ext}(0,\vec{\ell}\,\,)} 
\nonumber \\
&+2\log\prod_{{n',r\in \mathbb{Z}}\atop r>0}
\left|1-e^{2\pi i(rT+n'U+\vec{\ell}\odot\vec{V})}\right|^{\bar{c}_{\rm ext}(n'r,\vec{\ell}\,\,)}
+2\log\prod_{n=1}^\infty\left|1-e^{2\pi
i(nU+\vec{\ell}\odot{\cdot}\vec{V})}\right|^{\bar{c}_{\rm ext}(0,\vec{\ell}\,\,)}\bigg]
\nonumber \\
&+\bar{c}_{\rm ext}(0,\vec{0}\,\,)\left(\frac{\pi U_2}{3}-\ln Y+K\right)
+2\log\prod_{n=1}^\infty|1-e^{2\pi in U}|^{\bar{c}_{\rm ext}(0,\vec{0}\,\,)} \nonumber \\
&
+\frac{2U_2}{3\pi}+\frac{2\pi}{U_2}(\vec{\ell}\odot\Im\vec{V})
\left((\vec{\ell} \odot\Im\vec{V})+U_2\right)\ ,\label{TorusIntegralFinResult}
\end{align}
where $K=\gamma_E-1-\ln\frac{8\pi }{3\sqrt{3}}$, with $\gamma_E$ being the
Euler-Mascheroni constant.
Furthermore, we have introduced the shorthand notation for the modified
scalar-product
\begin{align}
\vec{\ell}\odot\vec{V}=\ell\cdot\Re\vec{V}+i\left|\vec{\ell}\cdot\Im\vec{V}
\right|\, .
\label{elldotV}
\end{align}
Here we have decided to work in a chamber of the moduli space where 
\begin{align}
\Im\vec{V}\in \left(\Lambda^+_{\mathfrak{e}_{8}}\oplus \Lambda^+_{\mathfrak{e}_{8}}\right) \otimes \mathbb{C}\ , \label{CondModSpaceSemi}
\end{align} 
such that the only contribution to the degenerate orbit with $\vec{\ell}\neq \vec{0}$ 
comes from vectors $\vec{\ell}=(\vec{\ell}_1,\vec{\ell}_2)$ where either $\vec{\ell}_1$ or $\vec{\ell}_2$ is 
a simple root of $\mathfrak{e}_8$ such that $\vec{\ell}\cdot \vec{\ell}=2$.

\subsection{Denominator Formula}\label{Sect:DenominatorFormula}

We shall now analyse the result (\ref{TorusIntegralFinResult}) in a little more detail. 
In the following we shall entirely focus on the logarithmic terms. Most of the non-logarithmic terms 
contribute to the Weyl vector $\rho$, appearing in the exponential prefactor of the  denominator 
formula (\ref{denominatorformula}) and ensure that the whole denominator formula has good modular properties under
$SL(2,\mathbb{Z})$ \cite{Borcherds1,Borcherds2}. Since these terms will not be of relevance for our present analysis we will 
suppress them in the following. 

The relevant part of (\ref{TorusIntegralFinResult}) can then be written as
\begin{align}
\mathcal{F}_1^{\text{analy}}(y)\sim \log||\Phi(y)||^2 +\cdots  \ ,
\end{align}
where we have defined
\begin{align}
\Phi(y)=\prod_{(r,n';\vec{\ell})>0} 
\left(1-e^{2\pi i (rT+n'U+\vec{\ell}\cdot\vec{V})}\right)^{\bar{c}_{\rm ext}(n'r-\vec{\ell}\cdot \vec{\ell}/2\,\,)}\ .
\label{LogTermInteg}
\end{align}
Furthermore, the range of the product  $(r,n';\vec{\ell}\,\,)>0$ is 
\begin{align}
&n'r-\frac{1}{2}\,\vec{\ell}\cdot \vec{\ell}\geq -1 &&\text{and}
&&\left\{\begin{array}{ll} r>0,\, 
n'\in \mathbb{Z},\,\vec{\ell}\in\Lambda_{\mathfrak{e}_8}\oplus\Lambda_{\mathfrak{e}_8} & \text{or} \\ 
r=0,\, n'> 0,\, \vec{\ell}\in\Lambda_{\mathfrak{e}_8}\oplus\Lambda_{\mathfrak{e}_8}  & \text{or} \\ 
r=n'=0,\, \vec{\ell}\in(\Lambda_{\mathfrak{e}_8}\oplus\Lambda_{\mathfrak{e}_8})^+ \ ,
\end{array}\right.\label{SumPosSimple}
\end{align} 
where $(\Lambda_{\mathfrak{e}_8}\oplus\Lambda_{\mathfrak{e}_8})^+$ denotes the positive part of the 
root lattice $\Lambda_{\mathfrak{e}_8}\oplus\Lambda_{\mathfrak{e}_8}$. 
The norm $||\cdot||^2$ in (\ref{LogTermInteg}) takes into account that there are contributions with 
$(r,n';\vec{\ell}\,\,)>0$ and contributions with $(r,n';\vec{\ell}\,\,)<0$. These conditions can be shown
to characterise the positive roots in $\Lambda^+_{\mathfrak{g}^{++}}$ with norm $2$ \cite{HPII}.

The key observation is now that we can identify\footnote{Alternatively, 
one can use the philosophy of  Borcherds-Gritsenko-Nikulin 
\cite{Borcherds1,Borcherds2,GritsenkoNikulin,Gritsenko:1996ax} to interpret (\ref{LogTermInteg}) 
as the denominator formula of an `automorphic correction' of $\mathfrak{g}^{++}$, where
$\mathfrak{g}=\mathfrak{e}_8 \oplus \mathfrak{e}_8$. This defines 
another Borcherds-Kac-Moody algbera $\mathcal{G}(\mathfrak{g}^{++})$, whose root multiplicities are 
directly defined by the coefficients $\bar{c}_{\text{ext}}$ in (\ref{cextbar}). This alternative point of view 
will be pursued in \cite{HPII}.} (\ref{LogTermInteg}) 
with a restriction of the denominator formula $\Phi_{\boru}(\hat{y}) $ for 
$\boru$ in (\ref{borudenominator}), where we set the weight vector $\vec{z}_3=0$ to zero and 
drop the terms that vanish in this limit, {\it i.e.}\ 
\begin{align}
\Phi(y)=\lim_{\vec{z}_3\rightarrow \vec{0}} \Phi_{\boru}(\hat{y})\big|_{\text{reg}}
=\prod'_{\alpha \in\Delta^+_{\boru}}\,  \left(1-e^{2\pi i (\alpha|y)}\right)^{c_{\text{ext}}(-\alpha^2/2)}\ .
\label{DenomFormPosRootwoWeyl}
\end{align}
Here the prime at the product means that we ignore the roots that lie entirely within the 
additional $\mathfrak{e}_8$ root lattice in (\ref{adde}) --- these give vanishing contributions since 
then $(\alpha|y)=0$. 

Thus $\mathcal{F}_1^{\text{analy}}$ is directly related to the 
denominator formula for the Borcherds algebra $\boru$ that was constructed explicitly in 
section~\ref{AlgebraBPSstates} using an auxiliary conformal field theory. The  restriction to 
$\vec{z}_3=0$ mirrors the fact that the additional 
root lattice in (\ref{adde}) was added by hand and does not play a role for the symmetry algebra
$\bor$ of the BPS spectrum.  The underlying physical reason for this
restriction is that we have been studying the problem in eight dimensions, {\it i.e.}\ we have
set some of the moduli (describing compactification along the additional directions) to 
special values. Thus we do not `see' the full root lattice of the underlying symmetry algebra. 
It would be interesting to understand in more detail the algebraic structure that arises
for amplitudes on $\mathbb{T}^n$ with $n>2$, see also section~\ref{Conclusions}.

\section{Discussion and Conclusions}\label{Conclusions}

In this paper we have constructed a Borcherds-Kac-Moody (BKM) algebra $\bor$ that 
acts on the perturbative BPS-states of the heterotic string theory on $\mathbb{T}^2$.
The Lie algebra $\bor$ plays in many ways the role of a `BPS-algebra', and its lattice of 
real roots coincides with that of the Lorentzian algebra 
$(\mathfrak{e}_8\oplus\mathfrak{e}_8)^{++}$.  We have shown that $\bor$, as well as
the closely related BKM algebra $\boru\supset \bor$, are relevant for the description 
of $\mathcal{N}=4$ threshold corrections.
More specifically, we have analysed a certain class of one-loop $\mathcal{N}=4$
topological amplitudes $\mathcal{F}_g$ in heterotic string theory compactified on 
$\mathbb{T}^6$. Upon splitting $\mathbb{T}^6=\mathbb{T}^4\times \mathbb{T}^2$ and 
taking the large volume limit of $\mathbb{T}^4$, we have shown that the analytic part of the 
simplest amplitude $\mathcal{F}_{g=1}^{\text{analy}}$ has an infinite product form which
can be identified with a certain restriction of the denominator formula for the BKM-algebra 
$\boru$. Furthermore, we have demonstrated that  the singularities of this amplitude are (partially) 
controlled by the Weyl group of $\bor$.

It would be interesting to extend the analysis beyond the case of
$\mathbb{T}^2$, and evaluate the integral (\ref{GenTopAmplitude}) for the full
Narain lattice of the six-torus $\mathbb{T}^6$.  For compactifications on
$\mathbb{T}^6$ the associated algebra would be constructed from an
indefinite Kac-Moody algebra of signature $(5, 21)$, for which the auxiliary CFT
has multiple temporal directions which complicates the description of its
physical states. Moreover, the Narain moduli space 
$SO(6,22)/(SO(6)\times SO(22))$ is no longer a hermitian symmetric domain and it 
is therefore unclear whether the theta correspondence affords an infinite product representation
which can be related to a denominator formula \cite{Borcherds2} (see also
\cite{Kiritsis:1997hf,Obers:1999um} for related discussions). 

Furthermore, we have neglected a detailed analysis of the
non-analytic part of $\mathcal{F}_{1}$, {\it i.e.}\ the part of the amplitude which
contributes to the `harmonic anomaly'. One of the main complications here is
that the result of the integral involves sums over polylogarithms of order $r>1$
(see {\it e.g.} \cite{Harvey:1995fq,Marino:1998pg}), which in particular cannot 
be written as infinite products. One might speculate that these terms can be
recast as an expansion in terms of characters of irreducible highest weight
representations of the BKM-algebra $\bor$.\footnote{Somewhat similar
speculations have been offered in the context of non-perturbative dyonic
BPS-states in \cite{Cheng:2008fc}.} To this end it might be useful to interpret
the full $\mathcal{F}_{1}$ as a `generalised prepotential', along the lines of
\cite{Kiritsis:1997hf}.   

Our results should have a dual type IIA interpretation, which ought to shed light
on the geometric meaning of the Fourier coefficients of the Jacobi forms in
(\ref{FourierExpansionConjugacy}). It is useful to look at the $\mathcal{N}=2$
situation for guidance. In this case, heterotic string theory on K3$\times
\mathbb{T}^2$ is dual to type II strings on a K3-fibered Calabi-Yau threefold
\cite{Aspinwall:1995vk}, and the Fourier coefficients of the modular forms
entering into the theta correspondence on the heterotic side become identified
with  the Gopakumar-Vafa invariants \cite{Gopakumar:1998ii,Gopakumar:1998jq} on
the type II side, thus explaining their integrality
\cite{Harvey:1995fq,Henningson:1996jf,Marino:1998pg,Klemm:2004km}. In the
$\mathcal{N}=4$ setting, on the other hand, heterotic string theory on
$\mathbb{T}^6$ is dual to type II string theory on K3$ \times \mathbb{T}^2$, and
it is therefore natural to speculate that the Fourier coefficients
(\ref{FourierExpansionConjugacy}), extended to the full amplitude
(\ref{GenTopAmplitude}) on $\mathbb{T}^6$,  are related to some topological
invariants of K3-surfaces. To this end, it would be necessary to generalise
the Gopakumar-Vafa analysis to the $\mathcal{N}=4$ situation, as discussed in
\cite{Katz:1999xq}. 

Finally, let us offer some further speculations as to the geometric role of the
$\mathcal{N}=4$ amplitudes $\mathcal{F}_{1}$. It is well-known that the one-loop
amplitude of the $\mathcal{N}=2$ B-model topological string on a Calabi-Yau
threefold can be written as a weighted product of Ray-Singer torsions
\cite{Bershadsky:1993cx}, thereby capturing information about the spectrum of
Laplacians on the complex structure moduli space of Calabi-Yau threefolds. It
would be interesting if a similar interpretation exists for the $\mathcal{N}=4$
amplitudes considered here, possibly related to determinants of Laplacians on
the moduli spaces of K3-surfaces as in \cite{Harvey:1996ts,JT}. We hope to
return to these and related issues in future work.

\vspace{.5cm}

\noindent{\bf \large{Acknowledgments}}

\vspace{.1cm}

\noindent We are grateful to Ignatios Antoniadis, Axel Kleinschmidt, Bengt E.W.
Nilsson, Christoffer Petersson, Boris Pioline, Yan Soibelman, Stephan
Stieberger and Roberto Volpato for helpful discussions and correspondences.
This work was partially supported by the Swiss National Science Foundation.

\appendix
\section{Infinite Dimensional Lie Algebras}\label{AppA}
\renewcommand{\theequation}{\Alph{section}.\arabic{equation}}
\setcounter{equation}{0}
In this appendix we shall describe some of the relevant background material
about infinite-dimensional
Lie algebras. We begin by discussing double extensions of (semi-)simple
finite Lie algebras,
which  are in particular Lorentzian subalgebras of Borcherds-Kac-Moody
algebras. We also review some general aspects of Borcherds-Kac-Moody algebras.

\subsection{Double Extensions of Finite Dimensional Lie Algebras}\label{App:DoubleExtension}

In this section we shall briefly sketch the affine and hyperbolic extensions of
finite-dimensional 
Lie algebras. We begin by discussing the case when the underlying finite dimensional
Lie algebra $\mathfrak{g}$ is simple and generalise later to semisimple Lie algebras. 

\subsubsection*{Extensions of Simple Lie
Algebras}\label{App:DoubleExtensionSimple}

Let $\Lambda_{\mathfrak g}$ be the root lattice of the finite-dimensional simple
Lie algebra $\mathfrak{g}$. 
A basis of $\Lambda_{\mathfrak g}$ is described by the positive simple roots
$\alpha_i$, $i=1,\ldots,r$,
where $r$ is the rank of the Lie algebra $\mathfrak{g}$. Every
finite-dimensional simple 
Lie algebra $\mathfrak{g}$  has a highest root  $\theta\in\Delta_+$ such that 
$\theta+\alpha_i \notin \Delta$ is not a root. 

We want to extend the root lattice of $\mathfrak{g}$ by a sublattice of
$\Pi^{1,1}$, the 
even unimodular lattice of dimension $2$. We denote the standard basis of
$\Pi^{1,1}$ by 
$\{\beta_1,\beta_2\}$, where we have the inner products 
\begin{equation}
(\beta_1|\beta_2)=1  \qquad 
(\beta_1|\beta_1)=(\beta_2|\beta_2)=0 \ ,\qquad 
(\beta_1|\alpha_i)=(\beta_2|\alpha_i)=0\quad \forall{i=1,\ldots,r}\ .
\end{equation}
Here the last identities mean that $\Pi^{1,1}$ is orthogonal to
$\Lambda_{\mathfrak g}$.

We now append the simple positive roots $\alpha_i$, $i=1,\ldots,r$ by the simple
roots 
(see \cite{Kac} and \cite{Gaberdiel:2002db}) 
\begin{align}
&\alpha_0=\beta_1-\theta &&\text{(affine root)}\\
&\alpha_{-1}=-\beta_1-\beta_2 &&\text{(hyperbolic root)}\ .
\end{align}
The inner product matrices of this new set of simple roots define the 
Cartan matrices of the affine extension $\mathfrak{g}^+$ and hyperbolic 
extension $\mathfrak{g}^{++}$ of $\mathfrak{g}$ respectively,
\begin{align}
&\mathcal{C}^{\mathfrak{g}^+}_{ab}=2\,\frac{(\alpha_a|\alpha_b)}{
(\alpha_a|\alpha_a)} &&a,b=0,\ldots,r \\[2pt]
&\mathcal{C}^{\mathfrak{g}^{++}}_{IJ}=2\,\frac{(\alpha_I|\alpha_J)}{
(\alpha_I|\alpha_J)} &&I,J=-1,\ldots,r\ .
\end{align}
The corresponding root lattices are given by
\begin{align}
&\Lambda_{\mathfrak{g}^{+}}=
\sum_{a=0}^{r}\mathbb{Z}\, \alpha_a\subset\Lambda_{\mathfrak{g}}\oplus\Pi^{1,1} &&\text{and} 
&&\Lambda_{\mathfrak{g}^{++}} =\sum_{I=-1}^{r}\mathbb{Z}\, \alpha_I =
\Lambda_{\mathfrak{g}}\oplus \Pi^{1,1}\ .
\end{align}
\subsubsection*{Extensions of Semisimple Lie
Algebras}\label{App:ExtendSemiSimple}
When the finite dimensional Lie algebra $\mathfrak{g}$ is semisimple, {\it i.e.}\ a direct
sum of simple Lie algebras, 
the extension procedure requires a little bit more care. Let $\mathfrak{g}$ be a
rank $r$ finite semisimple 
Lie algebra corresponding to a direct sum of $n$ simple subalgebras
\be
\mathfrak{g}=\bigoplus_{i=1}^{n} \mathfrak{g}_{(i)}.
\label{semisimple}
\ee
It follows that each $\mathfrak{g}_{(i)}$ is an ideal in $\mathfrak{g}$. 
Double extensions of semisimple Lie algebras have been discussed previously in 
\cite{Kleinschmidt:2008jj}, and we shall recall the salient features from
there.\footnote{Special 
cases of this construction were also considered earlier in
\cite{Henneaux:2006gp}.} As before, 
we begin by constructing the affine extension $\mathfrak{g}^{+}$. This is now
done in two steps. 
First we extend each individual summand of (\ref{semisimple}) into an affine
Kac-Moody algebra
\be 
\mathfrak{g}_{(i)}^{+}=\mathfrak{g}_{(i)}[[t,t^{-1}]]\oplus
\mathbb{C}c_{(i)}\oplus \mathbb{C}d_{(i)}\ ,
\ee
where $\mathfrak{g}_{(i)}[[t,t^{-1}]]$ is the loop algebra of
$\mathfrak{g}_{(i)}$ with spectral 
parameter $t$, $c$ is the central generator, and $d$ is the so called
`derivation'. We recall 
from \cite{Kac} that the derivation is needed in order to obtain a
non-degenerate inner product 
$(\cdot | \cdot )$ on the Cartan subalgebra of $\mathfrak{g}_{(i)}^{+}$. We
denote by 
${\tilde{\mathfrak{g}}^{+}}$ the direct sum of all $\mathfrak{g}_{(i)}^{+}$
\be
\tilde{\mathfrak{g}}^{+}=\bigoplus_{i=1}^{n}\mathfrak{g}_{(i)}^{+}\ ,
\ee
which is a again an affine Kac-Moody algebra. By extending each summand of 
(\ref{semisimple}) in this way, the resulting root lattice, which is a sublattice of 
$\Lambda_{\mathfrak{g}}\oplus \bigoplus_{i=1}^{n} \Pi_{(i)}^{1,1}$, 
is clearly too big; in fact, as in the simple case treated in section \ref{App:DoubleExtensionSimple}, 
we  are interested in constructing an affine extension $\mathfrak{g}^{+}$ whose root lattice 
$\Lambda_{\mathfrak{g}^{+}}$ is a sublattice of $\Lambda_{\mathfrak{g}}\oplus
\Pi^{1,1}$. 
To achieve this, we can now take the quotient of $\tilde{\mathfrak{g}}^{+}$ by
the 
$2(n-1)$-dimensional ideal generated by the elements 
$(c_{(1)}-c_{(2)}), (c_{(2)}-c_{(3)}), \dots, (c_{(n-1)}-c_{(n)})$ 
as well as the elements $(d_{(1)}-d_{(2)})$, 
$(d_{(2)}-d_{(3)}), \dots, (d_{(n-1)}-d_{(n)})$. We thus define 
the affine  extension of $\mathfrak{g}$ as 
\be
\begin{split}
\mathfrak{g}^{+}=&\tilde{\mathfrak{g}}^{+} /
\left(\bigoplus_{a=1}^{n-1}\mathbb{C} (c_{(a)}-c_{(a+1)})
\oplus \bigoplus_{a=1}^{n-1}\mathbb{C} (d_{(a)}-d_{(a+1)})\right)
\\
=&\mathfrak{g}[[t,t^{-1}]]\oplus \mathbb{C}c\oplus \mathbb{C}d \ ,
\end{split}
\label{semisimpleaffine}
\ee
where in the second line we have explicitly identified 
$c= c_{(1)}=\cdots =c_{(n)}$ as well as $d= d_{(1)}=\cdots =d_{(n)}$. 
Note that in contrast to $\tilde{\mathfrak{g}}^{+}$, the algebra $\mathfrak{g}^{+}$ 
is not a Kac-Moody algebra; it is however the physically relevant affine 
algebra in our context.\footnote{Although the physical context is different, 
the reasons for singling out the algebra 
$(\mathfrak{g}_{(1)}\oplus \cdots \oplus \mathfrak{g}_{(n)})^{+}$ 
are similar to the analysis in \cite{Kleinschmidt:2008jj}.}

The double extension $\mathfrak{g}^{++}$ is now obtained as before by promoting the 
derivation $d$ of (\ref{semisimpleaffine}) to a proper Cartan generator. The
structure of the resulting
algebra is most easily explained by first adding to $\tilde{\mathfrak{g}}^{+}$ a
new node that 
attaches with a  a single link to all the affine nodes of the individual
summands of $\tilde{\mathfrak{g}}^{+}$;
the resulting algebra will be denoted by $\tilde{\mathfrak{g}}^{++}$. The algebra
$\mathfrak{g}^{++}$ is then obtained
by dividing by a suitable ideal in $\tilde{\mathfrak{g}}^{++}$. To describe this
ideal, we observe that 
the  Cartan matrix associated with the Dynkin diagram of
$\tilde{\mathfrak{g}}^{++}$
is indefinite of rank $r+2$, and that it has one negative eigenvalue, as well as
$n-1$ zero eigenvalues (with
the remaining eigenvalues all being positive). Let us denote the null
eigenvectors as 
$u_a,\, a=1,\dots, n-1,$ with components 
$u_a^{(l)}$, $l=1,\dots, r+n+1$. We furthermore call the $r+n+1$ Cartan
generators of 
$\tilde{\mathfrak{g}}^{++}$ in the Chevalley basis $h_l$. It follows \cite{Kac}
that the center  
$\mathfrak{r}$ of $\tilde{\mathfrak{g}}^{++}$ is $(n-1)$-dimensional, and is
generated by the 
elements
\be
c_{a}=\sum_{l=1}^{r+n+1} u_a^{(l)} h_l\ .
\label{centersemisimpleextension}
\ee
The double extension $\mathfrak{g}^{++}$ may then be defined as the quotient of 
$\tilde{\mathfrak{g}}^{++}$ by the center 
\be
\mathfrak{g}^{++}=\tilde{\mathfrak{g}}^{++}\, /\, \mathfrak{r}\ . 
\label{semisimpledouble}
\ee
Again, we stress that the Lorentzian algebra so obtained is not a Kac-Moody algebra 
\cite{Kleinschmidt:2008jj}, but it is nevertheless the algebra that will be relevant 
in our context.

\subsection{Borcherds-Kac-Moody Algebras}
\label{BKMalgebras}

Next we want to give a very brief introduction to \emph{Borcherds-Kac-Moody} (BKM)
algebras that were first introduced in \cite{Borcherds0} (see also
\cite{Borcherds3,Borcherds4}).  These algebras are also sometimes referred to as 
Generalised Kac-Moody algebras or GKMs.

The BKM algebra $\bor$ is characterised 
by a Cartan matrix $\mathcal{C}$, which is now allowed to have infinite rank and
is generically of 
indefinite signature. Let $\{h_{\mathcal{I}}, e_\mathcal{I}, f_\mathcal{I}\}$,
$\mathcal{I}=1,\dots, \text{rank}\, \bor,$ be the set of Chevalley generators
subject to the relations 
(no summation on repeated indices)
\begin{equation}
\begin{array}{lll}
{}[h_{\mathcal{I}}, e_\mathcal{J}]=\mathcal{C}_{\mathcal{IJ}} \, e_\mathcal{J} \quad
& [h_{\mathcal{I}}, f_\mathcal{J}]=-\mathcal{C}_{\mathcal{IJ}} \, f_\mathcal{J} \quad
&[e_\mathcal{I}, f_\mathcal{J}]=h_\mathcal{IJ}\\
\text{ad}_{e_\mathcal{I}}^{1-\mathcal{C}_\mathcal{JI}}(e_\mathcal{J})=0
&\text{ad}_{f_\mathcal{I}}^{1-\mathcal{C}_\mathcal{JI}}(f_\mathcal{J})=0
&\forall\, \mathcal{C}_\mathcal{II}=2\ , \ \ \mathcal{I}\neq \mathcal{J} \\
{}[e_\mathcal{I},e_\mathcal{J}]=0
&[f_\mathcal{I},f_\mathcal{J}]=0
&\forall\,\mathcal{C}_{II}\leq 0\ , \ \ \mathcal{C}_\mathcal{JJ}<0\ ,\ \ \mathcal{C}_\mathcal{IJ}=0\ .
\end{array}
\end{equation}
As in the case of finite-dimensional  Lie algebras, all generators of  $\bor$
can be obtained by applying repeated commutators \cite{Kac,Borcherds0}. Furthermore,
the diagonal elements 
$h_\mathcal{I}$ generate the Cartan subalgebra $\mathcal{H}$, while the 
$e_\mathcal{I}$ and $f_\mathcal{I}$ generate nilpotent subalgebras 
$\mathcal{N}^{+}$ and $\mathcal{N}^{-}$, respectively. Thus, also BKM-algebras 
exhibit a standard triangular decomposition
\begin{align}
\bor =\mathcal{N}^{-}\oplus \mathcal{H}\oplus \mathcal{N}^{+}\ .
\end{align}
As for standard Kac-Moody algebras there is an invariant, non-degenerate
symmetric bilinear form
on $\mathcal{H}^{*}$, that we shall denote by $( \cdot | \cdot )$. However, the
main difference relative to standard Kac-Moody 
algebras is that the diagonal entries of this inner product are not required to
be positive. 
Thus the simple roots of $\bor$ come in two classes: \emph{real} simple roots
satisfying 
$(\alpha_\mathcal{I}|\alpha_\mathcal{I})>0$, and 
\emph{imaginary} simple roots satisfying
$(\alpha_\mathcal{I}|\alpha_\mathcal{I})\leq 0$. 

We denote by $\Delta$ the set of all roots. Generalising the usual terminology,
a root is said to be 
positive (resp.\ negative) if it is a non-negative (resp.\ non-positive) integer
linear combination 
of the simple roots. The set of roots thus splits again into a direct sum of
positive and 
negative roots, $\Delta=\Delta_+\oplus \Delta_-$. We also introduce the root
lattice $\Lambda_\mathcal{G}$ to be the integral span of all simple roots. It decomposes
as $\Lambda_\mathcal{G}=\Lambda_\mathcal{G}^{+}\cup \Lambda_\mathcal{G}^{-}$,
where $\Lambda_\mathcal{G}^{+}$ contains the non-negative integer linear
combinations 
of the simple roots, and similarly for $\Lambda_\mathcal{G}^{-}$.

The Weyl group $\mathcal{W}(\mathcal{G})$ is the group of reflections in 
$\Lambda_\mathcal{G}\otimes \mathbb{C}$ with respect to the real simple roots. 
In other words, upon denoting by $\alpha_I, \, I=1,\dots, n$ the real simple
roots, 
$\mathcal{W}(\mathcal{G})$ is generated by $n$ fundamental
reflections\footnote{Notice that 
$n$ need not be finite. For the construction of BKMs with an infinite number of
real simple 
roots see {\it e.g.}~\cite{Govindarajan:2010fu,Govindarajan:2008vi}.}
\be
w_I\ : \alpha\ \longmapsto \ \alpha
-2\frac{(\alpha|\alpha_I)}{(\alpha_I|\alpha_I)} \alpha_I, \qquad \quad
\alpha\in 
\Lambda_\mathcal{G}\otimes \mathbb{C}\ .
\ee
An additional important property of a BKM-algebra is the existence of a Weyl
vector $\rho$, satisfying 
\begin{align}
(\rho|\alpha)\leq -\frac{1}{2}(\alpha|\alpha)\ , \label{DefEquWeyl}  
\end{align}
with equality if and only if $\alpha$ is a simple root. For any (integrable)
lowest weight 
representation $R(\lambda)$ of $\mathcal{G}$ one further has the
Weyl-Kac-Borcherds 
character formula \cite{Kac,Borcherds0}
\be
\text{ch}\, R(\lambda)=\frac{\sum_{w\in\mathcal{W}} \epsilon(w)w(S)e^{-\rho}}
{\prod_{\alpha\in \Delta_+} (1-e^{\alpha})^{\text{mult}\, \alpha}}\ ,
\ee
where $\epsilon(w)=(-1)^{\ell(w)}$ with $\ell(w)$  the length of the Weyl
element $w$ (see {\it e.g.}\ \cite{Humphreys2}). This expression differs from the
standard Weyl-Kac character formula by the factor $w(S)$ which contains a
correction 
due to the imaginary simple roots \cite{Borcherds0}
\be
S=e^{\lambda+\rho}\sum_{\alpha\in \Lambda_\mathcal{G}^{+}} \xi(\alpha)e^{\alpha}
\ .
\ee
Here $\xi(\alpha)=(-1)^{m}$ if $\alpha$ is a sum of $m$ distinct pairwise
orthogonal 
imaginary simple roots which are orthogonal to $\lambda$, and $\xi(\alpha)=0$
otherwise. 
For our purposes we are interested in the simplest case of the trivial
representation $\lambda=0$, 
for which $\text{ch}\, R(\lambda)=1$, and the character formula reduces to the
so called 
\emph{denominator formula}
\be
\sum_{w\in \mathcal{W}} \epsilon(w)w(S)e^{-\rho}=
\prod_{\alpha\in \Delta_+}\left(1-e^{\alpha}\right)^{\text{mult}\, \alpha}\ .
\label{denominatorformula}
\ee
This formula relates a sum over the Weyl group $\mathcal{W}(\mathcal{G})$ to an
infinite 
product over all positive roots of $\mathcal{G}$. 


\section{One-Loop Integral in Terms of 
$SL(2,\mathbb{Z})$-Orbits}\label{App:Orbits}
\renewcommand{\theequation}{\Alph{section}.\arabic{equation}}
\setcounter{equation}{0}
In this appendix we will explicitly evaluate the three different contributions
in (\ref{IntegralSummaryRes}), corresponding to the different inequivalent $SL(2,{\mathbb Z})$ orbits. 

\subsection{The Zero Orbit}

The contribution from $A=0$ takes the form
\begin{align}
\mathcal{I}^{(\text{analy})}_{0}&=\int_{\mathbb{F}}\frac{d^2\tau}{\tau_2^2}
\frac{Y}{U_2}\sum_{{n\geq -1}\atop
{\vec{\ell}\in\Lambda_{\mathfrak{e}_8}\oplus\Lambda_{\mathfrak{e}_8}}}
\bar{c}_{\rm ext}(n,\vec{\ell}\,\, )\,\bar{q}^{\, n}
=\frac{2 i Y}{\pi U_2}\int_{\mathbb{F}}d^2\tau\,\frac{\partial}{\partial\tau}
\bigg[\sum_{{n\geq -1}\atop
{\vec{\ell}\in\Lambda_{\mathfrak{e}_8}\oplus\Lambda_{\mathfrak{e}_8}}}
\bar{c}_{\rm ext}(n,\vec{\ell}\,\,)\,\bar{q}^{\, n}\bigg]\ .
\end{align}
Performing an integration by parts and using modular invariance of the integrand
(see 
{\it e.g.}\ \cite{Lerche:1988np}), we can readily evaluate this integral to get
\begin{align}
\mathcal{I}^{(\text{analy})}_{0}&=\frac{2 Y}{\pi U_2}\lim_{\tau_2\to\infty}
\bigg[
\sum_{{n\geq -1}\atop {\vec{\ell}\in\Lambda_{\mathfrak{e}_8}\oplus\Lambda_{\mathfrak{e}_8}}}
\bar{c}_{\rm ext}(n,\vec{\ell}\,\,)\,\bar{q}^{\, n}\bigg]
=\frac{2\pi Y}{3U_2}\,\sum_{{\vec{\ell}\in
\Lambda_{\mathfrak{e}_8}\oplus\Lambda_{\mathfrak{e}_8}}}
\left[\bar{c}_{\rm ext}(0,\vec{\ell}\,\,)-24\bar{c}_{\rm ext}(-1,\vec{\ell}\,\,)\right]\ .
\end{align}

\subsection{The Non-Degenerate Orbit}
A representative matrix for the non-degenerate orbit can be taken to be 
\begin{equation}
A=\left(\begin{array}{cc}r & j \\ 0 & p\end{array}\right) \qquad \hbox{with}\qquad
\left\{\begin{array}{l}p\in\mathbb{Z}\neq 0 \\ r>j\geq 0 \ ,\end{array}\right.\label{NonDegMat}
\end{equation}
whereas the integration domain can be extended to the double-cover of the upper
half-plane
\begin{align}
\mathcal{I}_{ND}^{(\text{analy})}=\frac{2Y}{U_2}\int_{\mathbb{H}_+}
\frac{d^2\tau}{\tau_2^2}\,
\sum_{n\geq -1\atop \vec{\ell}\in\Lambda_{\mathfrak{e}_8}\oplus\Lambda_{\mathfrak{e}_8}}
\sum_{p\neq 0\atop r>j\geq 0}
\bar{c}_{\rm ext}(n,\vec{\ell}\,\,)\,\bar{q}^{\, n}\,
e^{2\pi i\vec{\ell}\cdot\vec{z}-\frac{\pi Y}{U_2^2\tau_2}\,|
\mathcal{A}|^2-2\pi ipr T+\frac{2\pi ir\Im(\vec{V})^2}{U_2^2}\,\mathcal{A}}\ .\nonumber
\end{align}
Performing the coordinate transformation
\begin{align}
&\tau'_1=-r\tau_1+j+pU_1&&\text{and}&&\begin{array}{l}\mathcal{A}
=\tau'_1+i(pU_2+r\tau_2) \\
\tilde{\mathcal{A}}=\tau'_1+i(-pU_2+r\tau_2)\end{array}
\end{align}
we find that all the $j$-dependence of the integrand is in the factor 
$\bar{q}^n=e^{-2\pi \tau_2n+\frac{2\pi in}{r}(\tau'_1-j-pU_1)}$ stemming from
the Fourier 
expansion (\ref{FourierExpansionConjugacy}). In this case, the summation over
$j$ yields 
only a non-vanishing result if $n$ is a multiple of $r$. We thus introduce
$n=n'r$, with
$n'\in\mathbb{Z}$ and obtain 
{\allowdisplaybreaks\begin{align}
\mathcal{I}_{ND}^{(\text{analy})}=&
-\frac{2Y}{U_2}\int_{\mathbb{H}_+}\frac{d^2\tau'}{\tau_2^2}
\sum_{{\vec{\ell}\in\Lambda_{\mathfrak{e}_8}\oplus\Lambda_{\mathfrak{e}_8}
\atop n',r\in \mathbb{Z}}\atop r>0}
\sum_{p\in \mathbb{Z}\atop p\neq 0}e^{-2\pi n'r\tau_2-2\pi in'(pU_1-\tau'_1)}
e^{2\pi i \vec{\ell}\cdot \vec{z}}\,\bar{c}_{\rm ext}(n'r,\vec{\ell}\,\,) \nonumber\\
&\times e^{-\frac{\pi Y}{U_2^2\tau_2}(\tau'^2_1+(pU_2+r\tau_2)^2)-2\pi iTrp
-\frac{2\pi i \Im (\vec{\ell}\cdot\vec{V})}{U_2}(\tau'_1-i\tau_2)-2\pi p \Re(\vec{\ell}\cdot \vec{V})}\ .
\end{align}}
The $\tau'_1$-integral is now Gaussian and can be solved using elementary methods,
leading to 
\begin{align}
\mathcal{I}_{ND}^{(\text{analy})}=&-2\sqrt{Y}\int_0^\infty
\frac{d\tau_2}{\tau_2^{3/2}}
\sum_{{\vec{\ell}\in\Lambda_{\mathfrak{e}_8}\oplus\Lambda_{\mathfrak{e}_8}\atop n',r,\in \mathbb{Z}}
\atop r>0}\sum_{p\in \mathbb{Z}\atop p\neq 0}e^{-2\pi n'r\tau_2
-2\pi in'pU_1-2\pi ip \Re(\vec{\ell}\cdot \vec{V})
-\frac{2\pi (\Im \vec{V})^2r}{U_2^2}(pU_2+r\tau_2)} \nonumber\\
&\times \bar{c}_{\rm ext}(n'r,\vec{\ell}\,\,)\,e^{-\frac{2\pi r \Im(\vec{\ell}\cdot \vec{V})\tau_2}{U_2}
-\frac{\pi Y}{U_2^2\tau_2}(pU_2+r\tau_2)^2
-2\pi iTrp-\frac{\pi\tau_2\left(n'U_2^2+(\Im \vec{V})^2r+U_2\Im (\vec{\ell}\cdot
\vec{V})\right)^2}{U_2^2  Y}}\,.
\end{align}
The integral over $\tau_2$ is of Bessel-type for which we can use the
identity
\begin{align}
&\int_0^\infty\frac{dx}{x^{3/2}}\,e^{-ax-b/x}=\sqrt{\frac{\pi}{b}}\
e^{-2\sqrt{ab}} &&\text{for}\,&&a>0\,\,\text{and }\,b>0\ .
\end{align}
In order to be able to use this relation, we need to specify the point in
moduli space at which we are working. Without loss of generality, we will assume
\begin{align}
&T_2>0 &&\text{and} &&U_2>0\ ,
\end{align}
in which case we only have to distinguish the cases (i) $\vec{\ell}\cdot (\Im\vec{V})>0$ 
and (ii) $\vec{\ell}\cdot (\Im\vec{V})<0$. We will in the following explicitly treat case 
(i), while case (ii) will follow similarly. Splitting 
the summation over $p$ into the pieces $p>0$ and $p<0$ we obtain
\begin{align}
\mathcal{I}_{ND}^{(\text{analy})}=&
\sum_{{\vec{\ell}\in\Lambda_{\mathfrak{e}_8}\oplus
\Lambda_{\mathfrak{e}_8}\atop n',r\in \mathbb{Z}}\atop r>0}
\sum_{p=1}^\infty\tfrac{2}{p}\, \left[e^{2\pi i p(rT+n'U+\vec{\ell}\cdot\vec{V})}
+e^{-2\pi i p(r\bar{T}+n'\bar{U}+\vec{\ell}\cdot\vec{\bar{V}})}\right]\,
\bar{c}_{\rm ext}(n'r,\vec{\ell}\,\,)\ .\label{OrbitnondegPre}
\end{align}
Recalling the identity $\sum_{l=1}^\infty\frac{x^l}{l}=\log(1-x)$, we can
perform the sum over $p$ to obtain the result
\begin{align}
\mathcal{I}_{ND}^{(\text{analy})}=2\log\prod_{{\vec{\ell}\in\Lambda_{\mathfrak{e}_8}\oplus\Lambda_{\mathfrak{e}_8}\atop 
n',r\in \mathbb{Z}}\atop r>0}\left|\ln(1-e^{2\pi i(rT+n'U+\vec{\ell}\cdot\vec{V})})
\right|^{\bar{c}_{\rm ext}(n'r,\vec{\ell}\,\,)}\ .
\end{align}
 Notice that the second term in (\ref{OrbitnondegPre}) is just the complex
conjugate of the first term, which explains the appearance of the absolute
square in the final result. The result for the case (ii), {\it i.e.}\ $\vec{\ell}\cdot (\Im\vec{V})<0$ 
can be obtained by replacing $(\Im\vec{V})$ by $-(\Im\vec{V})$.
\subsection{The Degenerate Orbit}
The last orbit to consider is the so-called degenerate orbit consisting of
matrices with vanishing determinant. We can pick a representative $A$ to be of
the form
\begin{align}
&A=\left(\begin{array}{cc}  0 &  j \\ 0 & p \end{array}\right) &&\text{with}
&&\left\{\begin{array}{l}(j,p)\neq (0,0) \\
j,p\in\mathbb{Z}\ , \end{array}\right.\label{DegenerateMatrix}
\end{align}
and we will integrate over the semi-infinite strip
$\mathbb{S}=\{\tau_1\in[-1/2,1/2],\tau_2\in[0,\infty)\}$. The integral then
becomes
\begin{align}
\mathcal{I}_{D}^{(\text{analy})}=\frac{Y}{U_2}
\int_{\mathbb{S}}\frac{d^2\tau}{\tau_2^2}&\sum_{\vec{\ell}\in
\Lambda_{\mathfrak{e}_8}\oplus\Lambda_{\mathfrak{e}_8}\atop n,j,p\in \mathbb{Z}}
\bar{c}_{\rm ext}(n,\vec{\ell}\,\,)\,\bar{q}^{\, n} e^{2\pi i \vec{\ell}\cdot \vec{z}
-\frac{\pi Y}{U_2^2\tau_2}|\mathcal{A}|^2}\ .
\end{align}
Notice that in this case the only $\tau_1$-dependence comes from the factor of
$\bar{q}^{\, n}$. The only non-vanishing contribution to the integral
$\tau_1\in[-1/2,1/2]$ therefore comes from $n=0$. For the remaining expression,
also the $\tau_2$ integration can be performed by elementary methods such that
we obtain
\begin{align}
\mathcal{I}_{D}^{(\text{analy})}&=\frac{U_2}{\pi}
\sum_{\vec{\ell} \in\Lambda_{\mathfrak{e}_8}\oplus\Lambda_{\mathfrak{e}_8}\atop (j,p)\neq (0,0)}\,
\frac{\bar{c}_{\rm ext}(0,\vec{\ell}\,\,)}{|j+p\, U|^2}\,\, e^{-\frac{2\pi i}{U_2} \vec{\ell}\cdot
\left[j(\Im\vec{V})+p\left(U_1(\Im\vec{V})-U-2(\Re\vec{V})\right)\right]}\ .
\end{align}
For the remaining sum over $j$ and $p$ it turns out to be useful to split the
summation into 
contributions for $\vec{\ell}=0$ and $\vec{\ell}\neq 0$. The former contribution
is identical to the 
one found in {\it e.g.}\ \cite{David:2006ji}, and equals\footnote{We are using here 
the same regularisation as in equation (B.19) of \cite{David:2006ji}.}
\begin{align}
\mathcal{I}^{(\text{analy})}_{D,\vec{\ell}=\vec{0}}&=\frac{U_2}{\pi}
\bar{c}_{\rm ext}(0,\vec{0}\,\,) \sum_{(j,p)\neq (0,0)}\frac{1}{|j+p\, U|^2} \\
&=\bar{c}_{\rm ext}(0,\vec{0}\,\,)\left(\frac{\pi U_2}{3}
-\ln Y+\gamma_E-1-\ln\frac{8\pi }{3\sqrt{3}}\right)
-\ln\prod_{n=1}^\infty|1-e^{2\pi in U}|^{4\bar{c}_{\rm ext}(0,\vec{0}\,\,)}\ .\nonumber
\end{align}
In order to calculate the contribution $\vec{\ell}\neq \vec{0}$ we first recall
that $\bar{c}_{\rm ext}(0,\vec{\ell}\,\,)=0$ for $\vec{\ell}\cdot\vec{\ell}> 2$. Moreover,
without loss of generality, we will assume to be working in a region in moduli
space where 
\begin{align}
&U_2>0 &&\text{and} &&U_2>\left|\vec{\ell}\cdot (\Im \vec{V})\right|\ .
\end{align}
In order to proceed, we have to distinguish two different contributions, namely
(i)~$\vec{\ell}\cdot (\Im\vec{V})> 0$ and (ii)~$\vec{\ell}\cdot (\Im\vec{V})<
0$. In the following we will explicitly work out the first case and indicate the
result for the second, which can be obtained in a similar fashion. With these
assumptions, we can write the sum over $j$ and $p$ as 
\begin{align}
\mathcal{I}_{D,\vec{\ell}\neq
0}^{(\text{analy})}=\frac{U_2}{\pi}&
\sum_{\vec{\ell}\in\Lambda_{\mathfrak{e}_8}\oplus\Lambda_{\mathfrak{e}_8}\atop 
\vec{\ell}\cdot\Im \vec{V}>\vec{0}}\bigg[\sum_{j\neq 0}\frac{\bar{c}_{\rm ext}(0,\vec{\ell}\,\,)}
{j^2}\, e^{-\frac{2\pi i}{U_2}\,\vec{\ell}\cdot(\Im\vec{V})\,j}\nonumber\\
&+\sum_{p,j\in\mathbb{Z}\atop p\neq 0}
\frac{\bar{c}_{\rm ext}(0,\vec{\ell}\,\,) }{(j+pU_1)^2+p^2U_2^2}\,e^{-\frac{2\pi i}{U_2}\,
\vec{\ell}\cdot(\Im\vec{V})\,j-\frac{2\pi i}{U_2}\,\vec{\ell}\cdot \left[U_1(\Im\vec{V}) 
-U_2(\Re\vec{V})\right]p}\bigg]\ .\label{DegOrbFirst}
\end{align}
In order to calculate these sums, we use the relations (see {\it e.g.}~\cite{David:2006ji})
\begin{align}
&\sum_{j=1}^\infty\frac{\cos\theta
j}{j^2}=\frac{\theta(\theta-2\pi)}{4}+\frac{\pi^2}{6}\,,\\
&\sum_{j=-\infty}^\infty\frac{e^{i\theta
j}}{(j+a_1)^2+a_2^2}=\frac{\pi}{a_2}\left[\frac{e^{-i\theta(a_1-ia_2)}}{1-e^{
-2\pi i(a_1-ia_2)}}+\frac{e^{-i\theta(a_1+ia_2)+2\pi i(a_1+ia_2)}}{1-e^{2\pi
i(a_1+ia_2)}}\right]\,,\label{SommerFeldWatson2}
\end{align}
where the second identity holds for $a_2>0$ and $0\leq \theta\leq 2\pi$. Using
the first identity in the first term in (\ref{DegOrbFirst}), we find explicitly
\begin{align}
\sum_{j\neq 0}\frac{e^{-\frac{2\pi i }{U_2} j\vec{\ell}\cdot
(\Im\vec{V})}}{j^2}=2\sum_{j=1}^\infty\frac{\cos\left(\frac{2\pi
}{U_2}j(\vec{\ell}\cdot
\Im\vec{V})\right)}{j^2}=\frac{2\pi^2}{U_2}\,(\vec{\ell}\cdot
\Im\vec{V})\left(\frac{(\vec{\ell}\cdot \Im\vec{V})}{U_2}-1
\right)+\frac{\pi^2}{3}\,.
\end{align}
Changing first $j\to -j$ and $p\to -p$ we can similarly treat the second term in
(\ref{DegOrbFirst}) using the second relation (\ref{SommerFeldWatson2}), and thus
obtain
\begin{align}
&\sum_{p,j\in\mathbb{Z}\atop p\neq 0}
\frac{e^{-\frac{2\pi i}{U_2}j(\vec{\ell}\cdot \Im\vec{V})}}{(j+p\, U_1)^2+p^2U_2^2}\,
e^{-\frac{2\pi i }{U_2}\, p\, \vec{\ell}\cdot \left(U_1 (\Im\vec{V})-U_2(\Re\vec{V})\right)} \\
&=\sum_{j=-\infty}^\infty\sum_{p=1}^\infty \left[\frac{e^{\frac{2\pi
i}{U_2}\left[j\vec{\ell}(\Im\vec{V})+p\,\vec{\ell}\cdot\left(U_1(\Im\vec{V})
-U_2(\Re\vec{V})\right)\right]}}{(j+p\, U_1)^2+p^2U_2^2}+\frac{e^{\frac{2\pi i}{U_2}
\left[j\vec{\ell}(\Im\vec{V})-p\,\vec{\ell}\cdot\left(U_1(\Im\vec{V})
-U_2(\Re\vec{V})\right)\right]}}{(j-p\, U_1)^2+p^2U_2^2}\right]\nonumber\\
&=\sum_{p=1}^\infty\frac{\pi}{p U_2}\bigg[\left(\frac{e^{-\frac{2\pi
i}{U_2}\,p\,\vec{\ell}\cdot (\Im\vec{V})\bar{U}}}{1-e^{-2\pi i
p\bar{U}}}+\frac{e^{-\frac{2\pi i}{U_2}\,p\,\vec{\ell}\cdot (\Im\vec{V})U+2\pi
ip U}}{1-e^{-2\pi i pU}}\right)e^{\frac{2\pi i }{U_2}p\vec{\ell}\cdot
\left(U_1(\Im\vec{V})-U_2(\Re\vec{V})\right)}\nonumber\\
&\hspace{1.9cm}+\left(\frac{e^{\frac{2\pi i}{U_2}\,p\,\vec{\ell}\cdot
(\Im\vec{V})U}}{1-e^{2\pi i pU}}+\frac{e^{\frac{2\pi i}{U_2}\,p\,\vec{\ell}\cdot
(\Im\vec{V})\bar{U}-2\pi ip\bar{U} }}{1-e^{-2\pi i
p\bar{U}}}\right)e^{-\frac{2\pi i }{U_2}p\vec{\ell}\cdot
\left(U_1(\Im\vec{V})-U_2(\Re\vec{V})\right)}\bigg]\ .\nonumber
\end{align}
Then we use the identity $\frac{1}{1-x}=\sum_{n=0}^\infty x^n$ to simplify this to 
\begin{align}
&\sum_{p=1}^\infty\sum_{n=0}^\infty\frac{\pi}{p \, U_2}\left[e^{2\pi i
p(nU+\vec{\ell}\cdot \vec{V})}+e^{2\pi ip(\vec{\ell}\cdot
\bar{\vec{V}}-(n+1)\bar{U})}+e^{-2\pi i p(n\bar{U}+\vec{\ell}\cdot
\bar{\vec{V}})}+e^{-2\pi ip(\vec{\ell}\cdot \vec{V}-(n+1)U)}\right]\nonumber\\
&=\sum_{p=1}^\infty \frac{\pi }{p \, U_2} \left[\sum_{n=1}^\infty \left(e^{2\pi i
p(nU+\vec{\ell}\cdot \vec{V})}+e^{2\pi ip(\vec{\ell}\cdot
\bar{\vec{V}}-n\bar{U})}\right)\right]+\text{c.c.}\nonumber\\
&=\frac{2\pi}{U_2}\left(\log\prod_{n=1}^\infty\left|1-e^{2\pi
i(nU+\vec{\ell}\cdot\vec{V})}\right|+\log\left|1-e^{2\pi
i\vec{\ell}\cdot\vec{V}}\right|\right)\ .
\label{expr}
\end{align}
\noindent 
Here we have used the explicit expression for the polylogarithms,
$\text{Li}_n(z)=\sum_{k=1}^\infty\frac{z^k}{k^n}$. Notice that the final
expression in (\ref{expr}) is manifestly real. The expression for
$\vec{\ell}\cdot (\Im\vec{V})<0$ can be obtained in exactly the same manner and
yields the same result with $(\Im\vec{V})$ replaced by $-(\Im\vec{V})$. The
total contribution of the degenerate orbit is therefore given by
\begin{align}
\mathcal{I}_{D}^{(\text{analy})}=& \bar{c}_{\rm ext}(0,\vec{0}\,\,)
\left(\frac{\pi U_2}{3}-\ln Y+\gamma_E-1-\ln\frac{8\pi}{3\sqrt{3}}\right)
-\ln\prod_{n=1}^\infty|1-e^{2\pi in U}|^{4\bar{c}_{\rm ext}(0,\vec{0}\,\,)} \nonumber\\
&+2\sum_{\vec{\ell}\in\Lambda_{\mathfrak{e}_8}\oplus
\Lambda_{\mathfrak{e}_8}\atop \vec{\ell}\neq\vec{0}}
\left(\log\prod_{n=1}^\infty
\left|1-e^{2\pi i(nU+\vec{\ell}\cdot\vec{V})}\right|^{\bar{c}_{\rm ext}(0,\vec{\ell}\,\, )}
+\log\left|1-e^{ 2\pi i\vec{\ell}\cdot\vec{V}}\right|^{\bar{c}_{\rm ext}(0,\vec{\ell}\,\,)}\right)
\nonumber\\
&+\frac{2U_2}{3\pi}+\frac{2\pi}{U_2}(\vec{\ell}\cdot\Im\vec{V})
\left[(\vec{\ell} \cdot\Im\vec{V})+U_2\right]\ .
\end{align}
Let us recall again that the contribution for $\vec{\ell}\cdot (\Im\vec{V})<0$
can be obtained by replacing $(\Im\vec{V})$ by $-(\Im\vec{V})$. Putting
everything together we then obtain (\ref{IntegralSummaryRes}).


\begin{thebibliography}{99}


\bibitem{Cecotti:1992rm}
S.~Cecotti and C.~Vafa,
{\it On classification of N=2 supersymmetric theories},
Commun.\ Math.\ Phys.\  {\bf 158} (1993) 569
\texttt{[arXiv:hep-th/9211097]}.
  
\bibitem{Seiberg:1994rs}
N.~Seiberg and E.~Witten,
{\it Monopole condensation, and confinement in N=2 supersymmetric Yang-Mills
theory},
Nucl.\ Phys.\  B {\bf 426} (1994) 19
(Erratum: ibid.\  B {\bf 430} (1994) 485)
\texttt{[arXiv:hep-th/9407087]}.

\bibitem{Sen:2007vb}
A.~Sen,
{\it Walls of marginal stability and dyon spectrum in N=4 supersymmetric string
theories},
JHEP {\bf 0705} (2007) 039
\texttt{[arXiv:hep-th/0702141]}.

\bibitem{Denef:2007vg}
F.~Denef and G.W.~Moore,
{\it Split states, entropy enigmas, holes and halos},
\texttt{arXiv: hep-th/0702146}.

\bibitem{Cheng:2008fc}
M.C.N.~Cheng and E.P.~Verlinde,
{\it Wall crossing, discrete attractor flow, and Borcherds algebra},
SIGMA {\bf 4} (2008) 068
\texttt{[arXiv:0806.2337 [hep-th]]}.
  
\bibitem{Gaiotto:2008cd}
D.~Gaiotto, G.W.~Moore and A.~Neitzke,
{\it Four-dimensional wall-crossing via three-dimensional field theory},
Commun.\ Math.\ Phys.\  {\bf 299} (2010) 163
\texttt{[arXiv:0807.4723 [hep-th]]}.
 
\bibitem{js}
D.~Joyce and Y.~Song, 
{\it A theory of generalized Donaldson-Thomas invariants},
\texttt{arXiv: 0810.5645 [math.AG]}.
  
\bibitem{ks}
M.~Kontsevich and Y.~Soibelman,
{\it Stability structures, motivic Donaldson-Thomas invariants and cluster
transformations},
\texttt{arXiv:0811.2435 [math.AG]}.

\bibitem{Alexandrov:2008gh}
S.~Alexandrov, B.~Pioline, F.~Saueressig and S.~Vandoren,
{\it D-instantons and twistors},
JHEP {\bf 0903} (2009) 044 
\texttt{[arXiv:0812.4219 [hep-th]]}.
  
\bibitem{Manschot:2010qz} J.~Manschot, B.~Pioline and A.~Sen, 
{\it Wall crossing from Boltzmann black hole halos,} 
\texttt{arXiv:1011.1258 [hep-th]}.
 
\bibitem{Harvey:1995fq} 
J.A.~Harvey and G.W.~Moore, 
{\it Algebras, BPS states, and strings}, 
Nucl.\ Phys.\  B {\bf 463} (1996) 315 
\texttt{[arXiv:hep-th/9510182]}.

\bibitem{Harvey:1996gc}
J.A.~Harvey and G.W.~Moore,
{\it On the algebras of BPS states},
Commun.\ Math.\ Phys.\  {\bf 197} (1998) 489
\texttt{[arXiv:hep-th/9609017]}.
  
\bibitem{Moore:1997ar}
G.W.~Moore,
{\it String duality, automorphic forms, and generalized Kac-Moody algebras},
Nucl.\ Phys.\ Proc.\ Suppl.\  {\bf 67} (1998) 56 
{\tt [arXiv:hep-th/9710198]}.

\bibitem{Borcherds0}
R.E.~Borcherds, 
{\it Generalized Kac-Moody algebras}, 
J.\ Algebra \textbf{115} (1988) 501.

 \bibitem{Gebert:1994mv}
R.W.~Gebert and H.~Nicolai,
{\it On E(10) and the DDF construction},
Commun.\ Math.\ Phys.\  {\bf 172} (1995) 571 
\texttt{[arXiv:hep-th/9406175]}.

\bibitem{Barwald:1997gm}
O.~Barwald, R.W.~Gebert, M.~Gunaydin and H.~Nicolai,
{\it Missing modules, the gnome Lie algebra, and E(10)},
Commun.\ Math.\ Phys.\  {\bf 195} (1998) 29
\texttt{[arXiv:hep-th/9703084]}.

\bibitem{Gebert:1997hx}
R.W.~Gebert and H.~Nicolai,
{\it On the imaginary simple roots of the Borcherds algebra g(II(9,1))},
Nucl.\ Phys.\  B {\bf 510} (1998) 721 
\texttt{[arXiv:hep-th/9705144]}.

\bibitem{HenryLabordere:2002dk}
P.~Henry-Labordere, B.~Julia and L.~Paulot,
{\it Borcherds symmetries in M-theory},
JHEP {\bf 0204} (2002) 049 
\texttt{[arXiv:hep-th/0203070]}.

\bibitem{HenryLabordere:2002xh}
P.~Henry-Labordere, B.~Julia and L.~Paulot,
{\it Real Borcherds superalgebras and M-theory},
JHEP {\bf 0304} (2003) 060 
\texttt{[arXiv:hep-th/0212346]}.

\bibitem{Henneaux:2010ys}
M.~Henneaux, B.L.~Julia and J.~Levie,
{\it $E_{11}$, Borcherds algebras and maximal supergravity},
\texttt{arXiv:1007.5241 [hep-th]}.

\bibitem{Borcherds1}
R.E.~Borcherds, 
{\it Automorphic forms on O(2,n) and infinite products}, 
Invent.\ Math.\ \textbf{120} (1995) 161.

\bibitem{Neumann:1997pr}
C.D.D.~Neumann,
{\it Perturbative BPS algebras in superstring theory},
Nucl.\ Phys.\  B {\bf 499} (1997) 596 
{\tt [arXiv:hep-th/9702197]}.

\bibitem{Fiol:2000wx}
B.~Fiol and M.~Marino,
{\it BPS states and algebras from quivers},
JHEP {\bf 0007} (2000) 031 
{\tt [arXiv:hep-th/0006189]}.

\bibitem{ks2}
M.~Kontsevich and Y.~Soibelman, 
{\it Cohomological Hall algebra, exponential Hodge structures and 
motivic Donaldson-Thomas invariants},
\texttt{arXiv:1006.2706 [math.AG]}.

\bibitem{Dijkgraaf:1996it}
R.~Dijkgraaf, E.P.~Verlinde and H.L.~Verlinde,
{\it Counting dyons in N=4 string theory},
Nucl.\ Phys.\  B {\bf 484} (1997) 543 
\texttt{[arXiv:hep-th/9607026]}.

\bibitem{GritsenkoNikulin}
V.A.~Gritsenko and V.V.~Nikulin, 
{\it Siegel automorphic form corrections to some Lorentzian Kac-Moody algebras},
C.\ R.\ Acad.\ Sci.\ Paris S\'er.\ A--B.\ {\bf 321} (1995) 1151.

\bibitem{Gritsenko:1996ax} 
V.A.~Gritsenko and V.V.~Nikulin, 
{\it The Igusa modular forms and `the simplest' Lorentzian Kac--Moody algebras},
{\tt arXiv:alg-geom/9603010}.

\bibitem{Cheng:2008gx}
M.C.N.~Cheng, {\it The spectra of supersymmetric states in string theory,}
\texttt{arXiv:0807. 3099 [hep-th]}.

\bibitem{Cheng:2008kt}
M.C.N.~Cheng and A.~Dabholkar,
{\it Borcherds-Kac-Moody symmetry of N=4 dyons},
Commun.\ Num.\ Theor.\ Phys.\  {\bf 3} (2009) 59
\texttt{[arXiv:0809.4258 [hep-th]]}.
  
\bibitem{Govindarajan:2008vi}
S.~Govindarajan and K.~Gopala Krishna,
{\it Generalized Kac-Moody algebras from CHL dyons},
JHEP {\bf 0904} (2009) 032 
\texttt{[arXiv:0807.4451 [hep-th]]}.

\bibitem{Govindarajan:2009qt}
S.~Govindarajan and K.~Gopala Krishna,
{\it BKM Lie superalgebras from dyon spectra in $\mathbb{Z}_N$ CHL orbifolds for
composite N},
JHEP {\bf 1005} (2010) 014 
\texttt{[arXiv:0907.1410 [hep-th]]}.

\bibitem{Govindarajan:2010fu} 
S.~Govindarajan, 
{\it BKM Lie superalgebras from counting twisted CHL dyons}, 
\texttt{arXiv: 1006.3472 [hep-th]}.

\bibitem{Antoniadis:2006mr} 
I.~Antoniadis, S.~Hohenegger and K.S.~Narain, 
{\it N = 4 topological amplitudes and string effective action}, 
Nucl.\ Phys.\  B {\bf 771} (2007) 40 
{\tt [arXiv:hep-th/0610258]}.

\bibitem{Borcherds2}
R.E.~Borcherds, 
{\it Automorphic forms with singularities on Grassmannians},
Invent.\ Math.\ {\bf 132} (1998) 491.

\bibitem{Antoniadis:2007cw} 
I.~Antoniadis, S.~Hohenegger, K.S.~Narain and E.~Sokatchev,
{\it Harmonicity in N=4 supersymmetry and its quantum anomaly}, 
Nucl.\ Phys.\  B {\bf 794} (2008) 348 
{\tt [arXiv:0708.0482 [hep-th]]}.

\bibitem{Berkovits:1994vy}
N.~Berkovits and C.~Vafa,
{\it N=4 topological strings},
Nucl.\ Phys.\  B {\bf 433} (1995) 123 
\texttt{[arXiv:hep-th/9407190]}.

\bibitem{Ooguri:1991fp} 
H.~Ooguri and C.~Vafa, 
{\it Geometry of N=2 strings},
Nucl.\ Phys.\  B {\bf 361} (1991) 469.

\bibitem{Antoniadis:2007ta} I.~Antoniadis and S.~Hohenegger, 
{\it Topological amplitudes and physical couplings in string theory}, 
Nucl.\ Phys.\ Proc.\ Suppl.\  {\bf 171 } (2007)  176
{\tt [arXiv:hep-th/0701290]}.

\bibitem{Bershadsky:1993ta}
M.~Bershadsky, S.~Cecotti, H.~Ooguri and C.~Vafa,
{\it Holomorphic anomalies in topological field theories},
Nucl.\ Phys.\  B {\bf 405} (1993) 279 
\texttt{[arXiv:hep-th/9302103]}.

\bibitem{Bershadsky:1993cx} 
M.~Bershadsky, S.~Cecotti, H.~Ooguri and C.~Vafa, 
{\it Kodaira-Spencer theory of gravity and exact results for quantum string
amplitudes}, 
Commun.\ Math.\ Phys.\  {\bf 165} (1994) 311 
{\tt [arXiv:hep-th/9309140]}.

\bibitem{Dixon:1990pc} 
L.J.~Dixon, V.~Kaplunovsky and J.~Louis, 
{\it Moduli dependence of string loop corrections to gauge coupling constants}, 
Nucl.\ Phys.\  B {\bf 355} (1991) 649.

\bibitem{Bump}
D.~Bump, 
{\it The Rankin-Selberg method: an introduction and survey},
{\texttt{http://sporadic.stanford.edu/bump/rallis.ps}}

\bibitem{Damour:2000hv}
T.~Damour and M.~Henneaux,
{\it E(10), BE(10) and arithmetical chaos in superstring cosmology},
Phys.\ Rev.\ Lett.\  {\bf 86} (2001)  4749
{\tt  [arXiv:hep-th/0012172]}.

\bibitem{Damour:2001sa}
T.~Damour, M.~Henneaux, B.~Julia and H.~Nicolai,
{\it Hyperbolic Kac-Moody algebras and chaos in Kaluza-Klein models},
Phys.\ Lett.\  B {\bf 509} (2001) 323 
\texttt{[arXiv: hep-th/0103094]}.

\bibitem{Damour:2002et}
T.~Damour, M.~Henneaux and H.~Nicolai, 
{\it Cosmological billiards,} Class.\ Quant.\ Grav.\  
{\bf 20} (2003)  R145
\texttt{[arXiv:hep-th/0212256]}.

\bibitem{Henneaux:2007ej}
M.~Henneaux, D.~Persson and P.~Spindel,
{\it Spacelike singularities and hidden symmetries of gravity},
Living Rev.\ Rel.\  {\bf 11} (2008) 1 
\texttt{[arXiv:0710.1818 [hep-th]]}.

\bibitem{Lerche:1999ju}
W.~Lerche and S.~Stieberger,
{\it 1/4 BPS states and non-perturbative couplings in N = 4 string theories},
Adv.\ Theor.\ Math.\ Phys.\  {\bf 3} (1999) 1539
{\tt  [arXiv:hep-th/9907133]}.

\bibitem{Bor0}
R.E.~Borcherds,
{\it Vertex algebras, Kac-Moody algebras and the Monster}, 
Proc.\ Nat.\ Acad.\ Sci.\ U.S.A.\ {\bf  83} (1986) 3068.

 \bibitem{EZ} 
M.~Eichler and D.~Zagier, 
{\it The Theory of Jacobi Forms}, 
Birkh\"auser (1985).

\bibitem{Antoniadis:1995zn} 
I.~Antoniadis, E.~Gava, K.S.~Narain and T.R.~Taylor, 
{\it N=2 type II heterotic duality and higher derivative F terms}, 
Nucl.\ Phys.\  B {\bf 455} (1995) 109 
{\tt [arXiv:hep-th/9507115]}.

\bibitem{Antoniadis:2010iq} 
I.~Antoniadis, S.~Hohenegger, K.S.~Narain and T.R.~Taylor, 
{\it Deformed topological partition function and Nekrasov backgrounds}, 
Nucl.\ Phys.\ B {\bf 838} (2010) 253
{\tt [arXiv:1003.2832 [hep-th]]}.

\bibitem{Marino:1998pg} 
M.~Marino and G.W.~Moore, 
{\it Counting higher genus curves in a Calabi-Yau manifold}, 
Nucl.\ Phys.\  B {\bf 543} (1999) 592 
{\tt [arXiv:hep-th/9808131]}.

\bibitem{Lerche:1987qk}
W.~Lerche, B.E.W.~Nilsson, A.N.~Schellekens and N.P.~Warner, 
{\it Anomaly cancelling terms from the elliptic genus,} 
Nucl.\ Phys.\  B {\bf 299} (1988) 91. 

\bibitem{Stieberger:1998yi} 
S.~Stieberger, 
{\it (0,2) heterotic gauge couplings and 
their M theory origin,} Nucl.\ Phys.\  B {\bf 541} (1999)  109
\texttt{[arXiv:hep-th/9807124]}.

\bibitem{Antoniadis:2009tr} I.~Antoniadis and S.~Hohenegger, 
{\it N=4 topological amplitudes and black hole entropy}, 
Nucl.\ Phys.\  B {\bf 837 } (2010) 61 {\tt [arXiv:0910.5596 [hep-th]]}.


\bibitem{Dijkgraaf}
R.~Dijkgraaf, {\it Mirror symmetry and elliptic curves},
in: ``The moduli space of curves'', Prog.\ Math.\ {\bf 129} (1995) 149.

\bibitem{KanekoZagier}
M.~Kaneko and D.~Zagier, 
{\it A generalized Jacobi theta function and quasi-modular forms},
in: ``The moduli space of curves'', Prog.\ Math.\ {\bf 129} (1995) 165.

\bibitem{LopesCardoso:1996nc} 
G.~Lopes Cardoso, G.~Curio and D.~Lust, 
{\it Perturbative couplings and modular forms in N = 2 string models with a
Wilson line}, 
Nucl.\ Phys.\  B {\bf 491} (1997) 147 
{\tt [arXiv:hep-th/9608154]}.

\bibitem{Obers:1999um}
N.A.~Obers and B.~Pioline,
{\it Eisenstein series and string thresholds},
Commun.\ Math.\ Phys.\  {\bf 209} (2000) 275
{\tt [arXiv:hep-th/9903113]}.

\bibitem{David:2006ud}
J.R.~David, D.P.~Jatkar and A.~Sen,
{\it Dyon spectrum in generic N = 4 supersymmetric Z(N) orbifolds},
JHEP {\bf 0701} (2007) 016
{\tt [arXiv:hep-th/0609109]}.

\bibitem{Wall} 
C.T.C.~Wall, 
{\it On the orthogonal groups of unimodular quadratic forms}, 
Math.\ Annalen {\bf 147} (1962) 328.

\bibitem{Banerjee:2007sr} 
S.~Banerjee and A.~Sen, 
{\it Duality orbits, dyon spectrum and gauge theory limit of heterotic string
theory on $T^6$},
JHEP {\bf 0803} (2008) 022 
{\tt [arXiv:0712.0043 [hep-th]]}.

\bibitem{LopesCardoso:1994ik}
G.~Lopes Cardoso, D.~Lust and T.~Mohaupt, 
{\it Threshold corrections and symmetry enhancement in string
compactifications},
Nucl.\ Phys.\  B {\bf 450} (1995) 115 
{\tt [arXiv: hep-th/9412209]}.

\bibitem{Foerger:1998kw} 
K.~Foerger and S.~Stieberger, 
{\it Higher derivative couplings and heterotic type I duality in eight-dimensions,} 
Nucl.\ Phys.\  B {\bf 559} (1999)  277 
\texttt{[arXiv:hep-th/9901020]}.

\bibitem{Lerche:1998nx} 
W.~Lerche and S.~Stieberger, 
{\it Prepotential, mirror map and F theory on K3,} 
Adv.\ Theor.\ Math.\ Phys.\  {\bf 2} (1998)  1105
\texttt{[arXiv:hep-th/9804176]}.

\bibitem{Kawai:1995hy}
T.~Kawai, 
{\it N=2 heterotic string threshold correction, K3 surface and generalized
Kac-Moody superalgebra,}
Phys.\ Lett.\  B {\bf 372} (1996)  59
\texttt{[arXiv:hep-th/9512046]}.

\bibitem{HPII}
S.~Hohenegger and D.~Persson, {\it in preparation}. 

\bibitem{Kiritsis:1997hf}
E.~Kiritsis and N.A.~Obers, 
{\it Heterotic/type-I duality in $D < 10$ dimensions, threshold corrections and D-instantons,} 
JHEP {\bf 9710} (1997) 004 
\texttt{[arXiv:hep-th/9709058]}.


\bibitem{Aspinwall:1995vk} 
P.S.~Aspinwall and J.~Louis, 
{\it On the ubiquity of K3 fibrations in string duality,} 
Phys.\ Lett.\  B {\bf 369} (1996) 233
\texttt{[arXiv:hep-th/9510234]}.
  
\bibitem{Gopakumar:1998ii} 
R.~Gopakumar and C.~Vafa, 
{\it M theory and topological strings. 1},
\texttt{arXiv:hep-th/ 9809187}.
 
\bibitem{Gopakumar:1998jq} 
R.~Gopakumar and C.~Vafa, 
{\it M theory and topological strings. 2}, 
\texttt{arXiv:hep-th/ 9812127}.
 
\bibitem{Henningson:1996jf} 
M.~Henningson and G.W.~Moore, 
{\it Counting curves with modular forms,} 
Nucl.\ Phys.\  B {\bf 472} (1996)  518
\texttt{[arXiv:hep-th/9602154]}.
  
\bibitem{Klemm:2004km} 
A.~Klemm, M.~Kreuzer, E.~Riegler and E.~Scheidegger, 
{\it Topological string amplitudes, complete intersection Calabi-Yau spaces and
threshold corrections,} 
JHEP {\bf 0505} (2005) 023 
\texttt{[arXiv:hep-th/0410018]}.

\bibitem{Kiritsis:2000zi}
E.~Kiritsis, N.A.~Obers and B.~Pioline, 
{\it Heterotic/type II triality and instantons on K3,} 
JHEP {\bf 0001} (2000) 029 
\texttt{[arXiv:hep-th/0001083]}.
\bibitem{Katz:1999xq} 
S.H.~Katz, A.~Klemm and C.~Vafa, 
{\it M theory, topological strings and spinning black holes,} 
Adv.\ Theor.\ Math.\ Phys.\  {\bf 3} (1999) 1445
\texttt{[arXiv:hep-th/9910181]}.

\bibitem{Harvey:1996ts} 
J.A.~Harvey and G.W.~Moore, 
{\it Exact gravitational threshold correction in the FHSV model,} 
Phys.\ Rev.\  D {\bf 57} (1998) 2329
\texttt{[arXiv:hep-th/9611176]}.
  
\bibitem{JT} 
J.~Jorgenson and A.~Todorov, 
{\it Enriques surfaces, analytic discriminants, and Bor\-cherds's $\Phi$-function,}
Commun. Math. Phys. {\bf 191} (1998) 249. 

\bibitem{Kac} V.G.~Kac, 
{\it Infinite-dimensional  Lie algebras}, 
Cambridge University Press (1990).

\bibitem{Gaberdiel:2002db}
M.R.~Gaberdiel, D.I.~Olive and P.C.~West,
{\it A class of Lorentzian Kac-Moody algebras},
Nucl.\ Phys.\  B {\bf 645} (2002) 403
{\tt [arXiv:hep-th/0205068]}.

\bibitem{Kleinschmidt:2008jj}
A.~Kleinschmidt and D.~Roest,
{\it Extended symmetries in supergravity: the semi-simple case},
JHEP {\bf 0807} (2008) 035
\texttt{[arXiv:0805.2573 [hep-th]]}.

\bibitem{Henneaux:2006gp}
M.~Henneaux, M.~Leston, D.~Persson and P.~Spindel,
{\it Geometric configurations, regular subalgebras of $E_{10}$ and M-theory
  cosmology},
JHEP {\bf 0610} (2006) 021
\texttt{[arXiv: hep-th/0606123]}.

\bibitem{Borcherds3} 
R.E.~Borcherds, 
{\it The Monster Lie algebra}, 
Adv.\ Math.\ \textbf{83} (1990) 30.

\bibitem{Borcherds4} 
R.E.~Borcherds, 
{\it Central extensions of generalized Kac-Moody algebras}, 
J.\ Algebra \textbf{140} (1991) 330.

\bibitem{Humphreys2}
J.E.~Humphreys, 
{\it Reflection groups and Coxeter groups},
Cambridge Studies in Advanced Mathematics, vol 29, 
Cambridge University Press (1990).

\bibitem{Lerche:1988np} 
W.~Lerche, A.N.~Schellekens and N.P.~Warner, 
{\it Lattices and strings,} 
Phys.\ Rept.\  {\bf 177} (1989) 1.

\bibitem{David:2006ji} 
J.R.~David, D.P.~Jatkar and A.~Sen,
{\it Product representation of dyon partition function in CHL models},
JHEP {\bf 0606} (2006) 064 
{\tt [arXiv:hep-th/0602254]}.
 

\end{thebibliography}
\end{document}